\documentclass[12pt,twoside,a4paper]{article}
\pdfoutput=1

\usepackage{fullpage}
\usepackage[english]{babel}
\usepackage[utf8x]{inputenc}
\usepackage{latexsym}
\usepackage[titletoc,toc,title,page]{appendix}
\usepackage{cite}
\usepackage{amsfonts}
\usepackage{amsmath}
\usepackage{amsthm}
\usepackage{amssymb}
\usepackage{graphicx}
\usepackage{mathtools}
\usepackage{enumerate}
\usepackage{mathrsfs}
\usepackage{verbatim}
\usepackage[toc]{appendix}
\usepackage{color}

\usepackage[width=\textwidth,font=small,labelfont=bf,
labelsep=endash]{caption}
\usepackage{titlesec}

\newcommand\Lame {Lam\'e\ }

\titleformat{\section}
  {\normalfont\fontsize{13}{16}\bfseries}{\thesection}{1em}{}
\titleformat{\subsection}
  {\normalfont\fontsize{11}{14}\bfseries}{\thesubsection}{1em}{}

\normalsize
\numberwithin{equation}{section}

\begin{document}

\title{\textbf{Dressed Minimal Surfaces in AdS$_4$}
\author{Dimitrios Katsinis$^{2,1}$, Dimitrios Manolopoulos$^1$, Ioannis Mitsoulas$^1$\\ and Georgios Pastras$^1$}
\date{\small $^1$ NCSR ``Demokritos'', Institute of Nuclear and Particle Physics,\\Agia Paraskevi 15310, Attiki, Greece\\
$^2$Department of Physics, National and Kapodistrian University of Athens,\\University Campus, Zografou, Athens 15784, Greece\linebreak \vspace{8pt}
\texttt{dkatsinis@phys.uoa.gr, manolopoulos@inp.demokritos.gr, mitsoulas@inp.demokritos.gr, pastras@inp.demokritos.gr}}}

\vskip .5cm

\maketitle

\begin{abstract}
We apply an arbitrary number of dressing transformations to a static minimal surface in AdS$_4$. Interestingly, a single dressing transformation, with the simplest dressing factor, interrelates the latter to solutions of the Euclidean non linear sigma model in dS$_3$. We present an expression for the area element of the dressed minimal surface in terms of that of the initial one and comment on the boundary region of the dressed surface. Finally, we apply the above formalism to the elliptic minimal surfaces and obtain new ones. \newline \newline \textbf{Keywords:} Minimal Surfaces, Integrable Systems, Dressing Method, Holographic Entanglement Entropy
\end{abstract}

\newpage

\tableofcontents

\newpage

\setcounter{equation}{0}
\section{Introduction}
\label{sec:introduction}

The gauge/gravity duality is a broad framework that relates the non-perturbative regime of a gauge theory to the weak coupling regime of a gravitational theory and vice versa. A concrete realization of the duality is provided by the AdS/CFT correspondence \cite{Maldacena:1997re,Gubser:1998bc,Witten:1998zw}, which states that $\mathcal{N}=4$ super Yang-Mills theory with $\mathrm{SU}(N)$ gauge group is equivalent to type-IIB superstring theory on AdS$_5\times$S$^5$ with $N$ units of flux through S$^5$. This setup has been studied extensively and a dictionary that imprints the identification of various quantities of the dual theories has been established (see for example \cite{Aharony:1999ti}). 

At the infinite t'Hooft coupling limit, the gravitational theory reduces to a classical one. At this limit, a prescription for the calculation of the holographic entanglement entropy was put forward by Ryu and Takayanagi \cite{Ryu:2006bv,Ryu:2006ef} and subsequently derived in the context of AdS/CFT in \cite{Lewkowycz:2013nqa}. The entanglement entropy is given by the von Neumann entropy associated with the \emph{reduced} density matrix that describes the degrees of freedom of a given subsystem. This subsystem is defined to contain the degrees of freedom in a given region of space, defined by a particular entangling surface. The calculation of entanglement entropy in quantum field theory is a formidable task, even for free field theories \cite{Srednicki:1993im,Callan:1994py,Holzhey:1994we,Calabrese:2004eu,Casini:2009sr}. The prescription of Ryu and Takayanagi states that the holographic entanglement entropy is proportional to the area of the co-dimension two minimal surface, which is anchored on the entangling surface at the boundary and extends towards the interior of the bulk. While this is a very well posed and clear prescription, in practise, its implementation is far from trivial, since one has to know the exact expression of the minimal surface in order to calculate its area. Even in the case of pure AdS geometries, very few minimal surfaces are known for an arbitrary number of dimensions, namely, minimal surfaces that correspond to spherical entangling surfaces or strip regions.

In $\textrm{AdS}_4$ there are extra tools that can be used compared to the general case. In this case, the co-dimension two minimal surfaces are two-dimensional Euclidean world-sheets. This implies that the minimal surface is a solution to the equations of motion derived from a Non-Linear Sigma Model (NLSM) action. In particular, the static co-dimension two minimal surfaces in AdS$_4$, which are the main subject of this work, are equivalent to co-dimension one minimal surfaces in the hyperbolic space H$^3$. Such two-dimensional Euclidean world-sheets, embedded in H$^d$, are of great interest, since they are the holographic duals of Wilson loops at strong coupling\cite{Rey:1998ik,Maldacena:1998im}. The general solution of the NLSM on H$^3$ was obtained in \cite{Ishizeki:2011bf} in terms of hyper-elliptic functions, while further aspects of it were studied in \cite{Kruczenski:2013bsa,Kruczenski:2014bla}. Key element of this solution, is the reducibility of the NLSMs defined on symmetric spaces, through the so called Pohlmeyer reduction \cite{Pohlmeyer:1975nb,Lund:1976ze}, to integrable equations of the family of the sine-Gordon equation. For a review on the subject see \cite{Miramontes:2008wt}. Given a solution of the Pohlmeyer reduced theory, the equations of motion of the NLSM become linear. The general solution was constructed by a clever incorporation of basic properties of hyper-elliptic functions. Yet, the practical use and qualitative understanding of this formal solution is very limited due to the high complexity of the hyper-elliptic functions. On a complementary approach in \cite{Pastras:2016vqu}, the whole class of solutions, whose Pohlmeyer field is expressed in terms of elliptic functions of only one of the two world-sheet coordinate, was derived through the ``inversion'' of the Pohlmeyer reduction and subsequently it was studied extensively.

Integrability has been extensively used in the context of AdS/CFT correspondance. For example, a basic aspect of the gauge/gravity duality concerns the relation of the spectra of  the dual theories. At the limit $N\rightarrow\infty$ and $\alpha^\prime\rightarrow 0$, with the t'Hooft coupling held fixed at a large value, the gauge theory is dual to non-interacting classical string theory. Exploiting the integrability properties of the theories on both sides of the duality, it was found that single trace operators of the gauge theory in the thermodynamic limit and classical string configurations were described by the same spectral curve. Thus, a formal matching of the spectra was achieved \cite{Beisert:2005bm,Beisert:2010jr}. It is interesting to investigate whether integrability can be used in a similar fashion in order to establish a direct relation between quantities relevant to entanglement entropy on the field theory side and its gravitational dual. The present study is a first attempt towards this direction. 

Expressing this kind of questions more concretely in field theory is beyond our understanding. The spectral curve, that corresponds to the solution \cite{Ishizeki:2011bf}, was constructed in \cite{Cooke:2014uga}. Yet, we lack any clue on how to relate entanglement entropy with a spectral curve. In this work we study some aspects of the dressing method \cite{Zakharov:1973pp,Zakharov:1980ty,Harnad:1983we,Hollowood:2009tw} on hyperbolic spaces and apply it on the elliptic minimal surfaces of \cite{Pastras:2016vqu} in order to construct new minimal surfaces. The dressing transformation can be perceived as an operation that changes the entangling surface and consequently the corresponding minimal surface. Obviously, this affects both the entanglement entropy in field theory, as well as the holographic entanglement entropy.

The dressing method is a technique that allows the construction of new NLSM solutions, once a solution is known, the so-called seed solution, by solving the auxiliary system, which is a system of first order partial differential equations. The implementation of the dressing method relies on the mapping of the solution of the NLSM to an element of an appropriate coset. There exist previous works that discuss the dressing of Wilson loops in AdS$_3$ and AdS$_5$ or AdS$_4\times $S$^2$ \footnote{As a matter of fact, in the latter the pseudoholomorphicity equations, which describe the Wilson loops as a result of supersymmetry, can effectively be described as a NLSM on S$^3$.}, using mappings on complex groups \cite{Jevicki:2007pk,Kalousios:2011hc}. The fact that the world-sheet metric is Euclidean causes complications to the construction of new real solutions. In these works, the problem is sidestepped, but this cannot be the case for arbitrary space-time dimensions. We apply the dressing method via the mapping of H$^3$ to the real coset $\mathrm{SO}(1,3)/\mathrm{SO}(3)$. We set up the problem from scratch and discuss in detail the constraints that have to be imposed on the solution of the auxiliary system. 

Contrary to most applications of the dressing method in the context of classical string solutions, such as \cite{Spradlin:2006wk,Kalousios:2006xy}, in the case of minimal surfaces, the Pohlmeyer reduced theory lacks a vacuum (either stable or unstable), and, thus, the simplest possible seeds are the elliptic minimal surfaces of \cite{Pastras:2016vqu}. As these seeds are highly non-trivial, more efficient techniques, such as the ones introduced in \cite{Katsinis:2018ewd}, are incorporated. Surprisingly, studying the dressing transformation of a general seed, we find that a single dressing transformation, with the simplest dressing factor, interrelates a real solution of the NLSM to a purely imaginary one. The imaginary solution of the Euclidean NLSM on hyperbolic space corresponds to a real solution of the Euclidean NLSM on de-Sitter space. This drawback leads us to study abstractly the dressing transformation for an arbitrary seed and to develop an iterative procedure that can be employed in order to construct new NLSM solutions once a solution of the auxiliary system is known. We discuss general quantitative aspects of the tower of solutions and present an algebraic addition formula for the surface element. Subsequently, we perform a double dressing transformation to the elliptic minimal surfaces. 

The rest of the paper is organised as follows: In section \ref{sec:Dressing} we discuss the dressing method for the Euclidean NLSM in $H^3$ for a general seed and an arbitrary number of dressing transformations, a relation between solutions of the NLSM on H$^3$ and solutions of the NLSM on dS$_3$ is established. In section \ref{sec:properties} we study some basic properties of the dressed surfaces, focusing on the transformation of the surface element and the entangling surface. In section \ref{sec:Dressed_Elliptic} we present the twice dressed elliptic minimal surfaces. Finally, in section \ref{sec:discussion} we discuss our results and possible future extensions.

\setcounter{equation}{0}
\section{Dressed Static Minimal Surfaces in AdS$_4$}
\label{sec:Dressing}

In view of the Ryu-Takayanagi prescription for the calculation of holographic entanglement entropy, the construction of a minimal surface for a given entangling surface presents interest not only from a mathematical point of view, but from a physical one as well. The main obstacle in finding minimal surfaces in an explicit form is the high complexity of the non-linear equations that govern them.

In AdS$_4$, co-dimension two minimal surfaces are two-dimensional, and, thus, they correspond to the special configurations, which extremize the Nambu-Gotto action, or equivalently a NLSM action, supplemented by the Virasoro constraints. We are interested in static minimal surfaces in AdS$_4$, which are equivalent to solutions of a Euclidean NLSM on the hyperbolic space H$^3$.

We consider the embedding of H$^3$ in the enhanced flat space $\mathbb{R}^{(1,3)}$, with coordinates $Y^0, Y^1, Y^2$ and $Y^3$. The H$^3$ submanifold is defined by the equation 
\begin{equation}
Y^T J Y \equiv - \left(Y^0\right)^2 + \left(Y^1\right)^2 + \left(Y^2\right)^2 + \left(Y^3\right)^2 = - \Lambda^2 ,
\label{eq:geom_constraint}
\end{equation}
where $J = \mathrm{diag} \left\{ -1,+1,+1,+1 \right\}$. In the following we set the scale of the hyperbolic space $\Lambda$ equal to one. Two-dimensional surfaces are parametrized by two real spacelike parameters $u$ and $v$. In corfomal gauge, the area of such a two-dimensional surface is given by the functional 
\begin{equation}
A=\int dzd\bar z \left( \left(\partial_+ Y \right)^T J \left( \partial_- Y \right) + \lambda \left(Y^T J Y+1\right)\right),
\end{equation}
where $z=(u+iv)/2$. We denote the associated derivatives as \footnote{We use the notation $\partial_\pm$ instead of the usual $\partial$ and $\bar{\partial}$ in order to have more compact expressions in what follows.}
\begin{equation}
\partial_+ \equiv \partial_{z}\qquad \partial _- \equiv \partial _{\bar z}.
\end{equation}
The coefficient $\lambda$ is a Lagrange multiplier, which enforces the geometric constraint \eqref{eq:geom_constraint}. The equations of motion assume the form 
\begin{equation}
\partial_+\partial_- Y=\left( \left(\partial_+ Y\right)^T J \left(\partial_- Y\right) \right)Y,
\end{equation}
while the Virasoro constraint reads
\begin{equation}
\left(\partial_+ Y\right)^T J \left(\partial_+ Y\right) = 0.
\end{equation}

The above equations can be reduced \`a la Pohlmeyer to the Euclidean $\cosh$-Gordon equation. Defining the Pohlmeyer field $a$ as
\begin{equation}
e^\alpha:=\left(\partial_+ Y\right)^T J \left(\partial_- Y\right),
\end{equation}
it can be shown that it obeys
\begin{equation}
\partial_+\partial_- \alpha=2\cosh\alpha.
\label{eq:cosh-gordon}
\end{equation}
The surface element is simply the exponential of the Pohlmeyer field, i.e. 
\begin{equation}
A=\int dzd\bar z e^\alpha.
\end{equation}

\subsection{The Dressing Method}

In a nutshell, the dressing method is a technique that enables one to construct a new solution of a NLSM given a known solution, the seed solution. The seed solution of the NLSM is mapped to an element $g$ of an appropriate coset, which is isomorphic to the symmetric target space of the NLSM. Then, instead of solving directly the second order non-linear equations of motion of the NLSM, one has to solve a pair of linear first order equations, the so called auxiliary system,
\begin{equation}
\partial_\pm\Psi(\lambda)=\frac{1}{1\pm\lambda}\left(\partial_\pm g\right)g^{-1}\Psi(\lambda),
\label{eq:aux_formal}
\end{equation}
where $\Psi(\lambda)$ is the auxiliary field, normilized as  $\Psi(0)=g$, and $\lambda$ is the spectral parameter. The equation of motion of the NLSM is the compatibility condition that must be obeyed, so that the auxiliary system \eqref{eq:aux_formal} has a solution.

A trivial gauge transformation of the auxiliary system $\Psi^\prime(\lambda)=\chi(\lambda)\Psi(\lambda),$ which is associated to a new solution of the NLSM. More details on the dressing method are provided in \cite{Spradlin:2006wk,Hollowood:2009tw}. As the NLSM that will occupy our interest, is defined with Euclidean world-sheet signature, there are a few, crucial, alterations with respect to the usual treatment of Lorentzian string world-sheets. 

\subsubsection{The Mapping between $\mathrm{H}^3$ and $\mathrm{SO}(1,3)/\mathrm{SO}(3)$}
In order to proceed with the dressing method, we need to establish the mapping between points of H$^3$ and elements of some appropriate coset, as was mentioned earlier. The hyperbolic space H$^3$ is isomorphic with the connected subspace of $\mathrm{SO}(1,3)/\mathrm{SO}(3)$, which contains the identity. The mapping of a vector of the enhanced space of H$^3$, namely $\mathbb{R}^{(1,3)}$, to an element $g$ of the coset $\mathrm{SO}(1,3)/\mathrm{SO}(3)$, which we use in the following is
\begin{equation}\label{eq:mapping}
g=\left(I+2Y_0 Y_0^TJ\right)\left(I+2Y Y^TJ\right),
\end{equation}
where $I$ is the identity matrix, $J=\mathrm{diag}\{-1,1,1,1\}$ is the metric of the enhanced space and $Y_0$ is a constant vector of H$^3$, i.e. $ Y_0^T J Y_0= - 1$. We denote
\begin{equation}
\theta:=I+2Y_0 Y_0^TJ.
\end{equation}
The special choice 
\begin{equation}
Y_0^T=\begin{pmatrix}
1 & 0 & 0& 0
\end{pmatrix}
\label{eq:special_representative}
\end{equation}
corresponds to $\theta=J$. 

It can be easily shown that the element $g$, given by \eqref{eq:mapping}, possesses the following properties
\begin{align}
\bar{g} &= g ,\\
\theta g \theta g&=I,\label{eq:coset_con}\\
g^TJg&=J,\label{eq:inv_con}
\end{align}
which state that $g$ is an element of the coset $\mathrm{SO}(1,3)/\mathrm{SO}(3)$.

\subsubsection{Constraints}
\label{subsec:constraints}
In the following, we derive the appropriate constraints, which ensure that the dressed solution $g^\prime=\chi(0)g$ is also an element of the coset $\mathrm{SO}(1,3)/\mathrm{SO}(3)$, as consistency conditions of the solution of the auxiliary system. In doing so, we consider a general constant matrix $\theta$ and not the special choice \eqref{eq:special_representative}. The analysis draws heavily on \cite{Combes:1993rw}. Since we work with a Euclidean NLSM, the main difference to the case of dressed string solutions is related to the constraint imposed by complex conjugation. 

We set $\lambda\rightarrow\bar{\lambda}$ in the auxiliarry system \eqref{eq:aux_formal} and consider the complex conjugate of these equations
\begin{equation}\label{eq:aux_formal_complex}
\partial_\mp\bar{\Psi}(z,\bar{z};\bar{\lambda})=\frac{1}{1\pm\lambda}\left(\partial_\mp g\right)g^{-1}\bar{\Psi}(z,\bar{z};\bar{\lambda}).
\end{equation}
Clearly, the two pairs of equations \eqref{eq:aux_formal} and \eqref{eq:aux_formal_complex} are compatible only if
\begin{equation}\label{eq:reality_constraint}
\bar{\Psi}(\bar{\lambda})= \Psi(-\lambda)m_1(\lambda),
\end{equation}
where $m_1(\lambda)$ is an arbitrary constant matrix which obeys $m_1(\lambda)\bar{m}_1(-\bar{\lambda})=I$ \footnote{This is required, since acting twice with complex conjugation should result in $\Psi$.}. The constraint \eqref{eq:reality_constraint} is general, in the sense, that any auxiliary system, defined on a real coset, with Euclidean world-sheet coordinates must obey it.

Next, we set $\lambda\rightarrow 1/\lambda$ into \eqref{eq:aux_formal}. Furthermore, equation \eqref{eq:coset_con} implies $(\partial_\pm g)\theta g+g\theta (\partial_\pm g)=0$, and, thus,
\begin{equation}
\partial_\pm\left[g\theta\Psi(1/\lambda)\theta\right]=\frac{1}{1\pm\lambda}\left(\partial_\pm g\right)g^{-1}\left[g\theta\Psi(1/\lambda)\theta\right].
\end{equation}
Consequently, 
\begin{equation}\label{eq:coset_constraint}
g\theta\Psi(1/\lambda)\theta=\Psi(\lambda)m_2(\lambda),
\end{equation}
where $m_2(\lambda)$ is an arbitrary constant matrix which obeys $m_2(\lambda)\theta m_2(1/\lambda)\theta=I$ \footnote{This constraint ensures that performing the transformation $\lambda\rightarrow 1/\lambda$ twice results the in the initial matrix $\Psi$.}.

Finally, from \eqref{eq:inv_con}, it follows that $J\left[\left(\partial_\pm g\right)g^{-1}\right]^TJ=-\left(\partial_\pm g\right)g^{-1}$. Thus,
\begin{equation}
\partial_\pm\left[J\Psi(\lambda)^TJ\right]^{-1}=\frac{1}{1\pm\lambda}\left(\partial_\pm g\right)g^{-1}\left[J\Psi(\lambda)^TJ\right]^{-1},
\end{equation}
which implies that
\begin{equation}\label{eq:subgroup_constraint}
\left[J\Psi(\lambda)^TJ\right]^{-1}=\Psi(\lambda)m_3(\lambda),
\end{equation}
where the matrix $m_3(\lambda)$ must obey $J m_3^T(\lambda) J = m_3(\lambda)$.

To sum up, the fact that the element $g$ belongs to the coset $\mathrm{SO}(1,3)/\mathrm{SO}(3)$ implies the constraints \eqref{eq:reality_constraint}, \eqref{eq:coset_constraint} and \eqref{eq:subgroup_constraint} on the solution of the auxiliary system.

\subsubsection{The Dressing Factor}
In this section we will construct the simplest dressing factor $\chi\left(\lambda\right)$, following \cite{Hollowood:2009tw}. More general ones can be constructed using the results of \cite{Harnad:1983we}. We will discuss them subsequently.

Demanding that the dressed auxiliary field solution, $\Psi^\prime$ obeys the constraints \eqref{eq:reality_constraint}, \eqref{eq:coset_constraint} and \eqref{eq:subgroup_constraint}, as $\Psi$ does, so that the dressed element $g^\prime$ also belongs to the coset $\mathrm{SO}(1,3)/\mathrm{SO}(3)$, implies that the dressing factor must obey the following constraints:
\begin{align}
\bar{\chi}\left(\bar{\lambda}\right)&=\chi\left(-\lambda\right),\label{eq:chi_reality}\\
\chi\left(1/\lambda\right)&=g^\prime J \chi\left(\lambda\right)gJ,\label{eq:chi_coset}\\
\chi^{-1}(\lambda)&=J\chi^T(\lambda)J.\label{eq:chi_inverse}
\end{align}
We have assumed that the matrices $m_1$, $m_2$ and $m_3$ are the same for the seed and dressed solutions. Without loss of generality, we choose $m_1 = I$, $m_2 = - J$ and $m_3 = I$ in what follows.

In general, the dressing factor is a meromorphic function of $\lambda$, and, thus, has an expansion of the form
\begin{equation}
\chi\left(\lambda\right)=I+ \sum_i\frac{Q_i}{\lambda-\lambda_i}.
\end{equation}
The constraints \eqref{eq:chi_reality}, \eqref{eq:chi_coset} and \eqref{eq:chi_inverse} enforce the poles in this expression to come in quadruplets of the form $\left\{ \lambda_i , - \bar{\lambda_i} , \lambda_i^{-1} , -\bar{\lambda}_i^{-1} \right\}$. Naively, it follows that the simplest dressing factor has the following structure
\begin{equation}
\chi\left(\lambda\right)=I+\frac{Q}{\lambda-\lambda_1}-\frac{\bar{Q}}{\lambda+\bar{\lambda_1}}+\frac{\tilde{Q}}{\lambda-\lambda_1^{-1}}-\frac{\bar{\tilde{Q}}}{\lambda+\bar{\lambda_1}^{-1}},
\end{equation}
while the inverse of the dressing factor can be obtained by \eqref{eq:chi_inverse}. In addition, this form of $\chi$ ensures that the constraint \eqref{eq:chi_reality} is satisfied. Then, equating the residues of the left-hand-side and the right-hand-side of \eqref{eq:chi_coset} we obtain
\begin{equation}\label{eq:chi_coset_pole}
\tilde{Q}=-\frac{1}{\lambda_1^2}g^\prime JQgJ,
\end{equation}
while the analytic part of \eqref{eq:chi_coset} implies that
\begin{equation}\label{eq:chi_coset_analytical}
\chi(0)g J\chi(0)=g J.
\end{equation}
Finally, the equations of motion of the dressing factor read
\begin{equation}\label{eq:chi_eom}
\left(1\pm\lambda\right)\left(\partial_\pm \chi\right)\chi^{-1}+\chi\left(\partial_\pm g\right)g^{-1}\chi^{-1}=\left(\partial_\pm g^\prime\right)g^{\prime-1}.
\end{equation}
For $\lambda=0$ these equations are satisfied trivially, thus one needs only to ensure that the residues of the various poles cancel.

The most economical way to satisfy the constraints is by choosing the poles to lie on the imaginary axis, i.e. demanding
\begin{equation}
\lambda_1= i \mu_1,
\end{equation}
where $\mu_1\in\mathbb{R}.$ This implies that the locations of the poles at $\lambda_1$ and $-\bar{\lambda_1}$ coincide. After appropriate redefinitions, the dressing factor is expressed as
\begin{equation}\label{eq:chi_2_poles}
\chi=I + i\frac{\mu_1+\mu_1^{-1}}{\lambda-i\mu_1}Q-i \frac{\mu_1+\mu_1^{-1}}{\lambda+i\mu_1^{-1}}\tilde{Q},
\end{equation}
where $\bar{Q}=Q$ and $\bar{\tilde{Q}}=\tilde{Q}.$ The inverse of the dressing factor can be obtained using \eqref{eq:chi_inverse}. Moreover, the above expression satisfies the constraint \eqref{eq:chi_reality}. For convenience, we will specify the appropriate relation between $Q$ and $\tilde{Q}$, which is necessary for the satisfaction of the constraint \eqref{eq:chi_coset}, later. Next, we impose the relation $\chi\chi^{-1}=I.$ The cancellation of the residues of the first order poles at $i\mu_1$ and $-i\mu_1^{-1}$ implies that
\begin{align}
Q\left(I -J\tilde{Q}^TJ\right)+\left(I -\tilde{Q}\right)JQ^TJ&=0,\\
\tilde{Q}\left(I -JQ^TJ\right)+\left(I -Q\right)J\tilde{Q}^TJ&=0.
\end{align}
Clearly, both relations are satisfied if
\begin{equation}
\tilde{Q}=JQ^TJ \label{eq:inversion_Q_Qtilde}
\end{equation}
and $Q$ is a projection matrix, i.e. it satisfies $Q^2=Q$. 
The cancellation of the residues of the second order poles at the same locations requires that
\begin{align}
Q^TJQ=QJQ^T&=0,\label{eq:inversion_second_order}\\
\tilde{Q}^TJ\tilde{Q}=\tilde{Q}J\tilde{Q}^T&=0.\label{eq:inversion_second_order_tilde}
\end{align}
The equation \eqref{eq:inversion_second_order_tilde} is redundant, as it follows from equations \eqref{eq:inversion_Q_Qtilde} and \eqref{eq:inversion_second_order}. Furthermore, these two equations imply that
\begin{equation}
Q \tilde{Q} = \tilde{Q} Q = 0. \label{eq:inversion_prod_QQtilde}
\end{equation}

We parametrize the matrix $Q$ as
\begin{equation}
Q=\frac{J H F^T}{F^T JH},
\end{equation}
where $F$ and $H$ are real vectors. Then, equation \eqref{eq:inversion_Q_Qtilde} implies that
\begin{equation}
\tilde{Q}=\frac{J F H^T}{F^T JH}.
\end{equation}
The constraints \eqref{eq:inversion_second_order} suggest that
\begin{equation}
H^TJH=F^TJF=0.
\end{equation}

Returning now to the equations of motion, the right-hand-side of \eqref{eq:chi_eom} does not depend on $\lambda,$ thus, the same must hold for the left-hand-side. The cancellation of the residues of the second order poles at $i\mu_1$ and $-i\mu_1^{-1}$ suggests
\begin{align}
\left(1\pm i \mu_1\right)&\partial_\pm F^T+F^T\left(\partial_\pm g\right)g^{-1}=0,\\
\left(1\mp i \mu_1^{-1}\right)&\partial_\pm H^T+H^T\left(\partial_\pm g\right)g^{-1}=0.
\end{align}
These equations imply that
\begin{align}
F^T&=p^T J \Psi^{-1}(i\mu_1),\label{eq:eom_F}\\
H^T&=p^T J \Psi^{-1}(-i\mu_1^{-1})\label{eq:eom_H},
\end{align}
where $p$ is a constant vector. We remind the reader that $\Psi(\lambda)$ is real whenever $\lambda$ is purely imaginary as a consequence of equation \eqref{eq:reality_constraint}. Moreover, the vectors $F$ and $H$ obey that
\begin{equation}
gH=-JF,
\end{equation}
in virtue of \eqref{eq:coset_constraint}. This relation implies
\begin{equation}
\tilde{Q}=gJ Q gJ. \label{eq:inversion_gJQjG}
\end{equation}

We have not yet enforced that the dressing factor with only two poles \eqref{eq:chi_2_poles} satisfies the constraint \eqref{eq:chi_coset}. For the generic four-pole dressing factor, this constraint results in equations \eqref{eq:chi_coset_pole} and \eqref{eq:chi_coset_analytical}. In the case of the two-pole dressing factor \eqref{eq:chi_2_poles}, the first one reads
\begin{equation}
\tilde{Q}=-\frac{1}{\mu_1^2}g^\prime JQgJ,
\end{equation}
It is simple to show that this relation, as well as \eqref{eq:chi_coset_analytical}, are indeed satisfied, as a consequence of equations \eqref{eq:inversion_prod_QQtilde}, \eqref{eq:inversion_gJQjG} and the fact that $Q$ is a projective operator.

Equation \eqref{eq:inversion_second_order} holds if the vector $p$ obeys
\begin{equation}
p^TJp=0.
\end{equation}
In addition, both $Q$ and $\tilde{Q}$ are real, as required, provided that $p=\bar{p}$. Finally, it is a matter of algebra to show that the residues of the first order poles of the equations of motion cancel as long as \eqref{eq:eom_F} and \eqref{eq:eom_H} hold, thus the equations of motion are satisfied.

To sum up, the simplest dressing factor reads
\begin{equation}\label{eq:definition_chi_simplest_W}
\chi(\lambda)=I+i\frac{\mu_1+\mu_1^{-1}}{\lambda-i\mu_1}g\frac{J W W^T J}{W^T g^{-1}W}-i\frac{\mu_1+\mu_1^{-1}}{\lambda+i\mu_1^{-1}}\frac{W W^T}{W^Tg^{-1}W}g^{-1},
\end{equation}
where
\begin{equation} \label{eq:W_definition}
W=\Psi(i\mu_1)p .
\end{equation}
The vector $W$ is null, i.e. $W^T J W = 0$.

Using \eqref{eq:mapping}, it is straightforward to show that the dressed element of the coset reads
\begin{equation}
g^\prime = J - 2J\left(\frac{Y}{\mu_1}+\frac{\mu_1+\mu_1^{-1}}{2}\frac{JW}{W^TY}\right)\left(\frac{Y}{\mu_1}+\frac{\mu_1+\mu_1^{-1}}{2}\frac{JW}{W^TY}\right)^TJ,
\end{equation}
which implies that the dressed solution of the NLSM, expressed as a vector in the enhanced space of H$^3$, is
\begin{equation}\label{eq:Y_prime}
Y^\prime = i\left(\frac{Y}{\mu_1}+\frac{\mu_1+\mu_1^{-1}}{2}\frac{JW}{W^TY}\right).
\end{equation}
The vector $Y^\prime$ satisfies the equations of motion and the Virasoro constraints, nevertheless it is purely imaginary. The imaginary part of this vector satisfies the equations of motion of the Euclidean NLSM defined on dS$_3$ and not in H$^3$. Expecting that the converse is also true, we apply an arbitrary number of dressing transformations in an iterative fashion in order to obtain new real solutions, whenever this number is even.

\subsection{Multiple Dressing}
\label{subsec:multiple_dressing}
Let $g_0$ be the original seed solution. Via a single dressing transformation we construct a dressed solution $g_1$. This in turn may play the role of the seed solution for another transformation. Pictorially,
\begin{equation}
g_0 \xrightarrow{\scriptsize \chi_1(0)} g_1\xrightarrow{\scriptsize\chi_2(0)} g_2\ldots g_{k-1}\xrightarrow{\scriptsize\chi_{k}(0)} g_{k}.
\end{equation}
Let $\Psi_k(\lambda)$ denotes the solution of the auxiliary system which incorporates the solution $g_{k-1}$ as the seed solution, namely
\begin{equation}\label{eq:psi_k_auxliary}
\partial_\pm\Psi_k(\lambda)=\frac{1}{1\pm\lambda}\left(\partial_\pm g_{k-1}\right)g_{k-1}^{-1}\Psi_k(\lambda) , \quad g_{k-1} = \Psi_k \left( 0 \right) .
\end{equation}
Then, in an obvious manner,
\begin{equation}\label{eq:psi_k}
\Psi_{k}(\lambda)=\chi_{k-1}(\lambda)\Psi_{k-1}(\lambda).
\end{equation} 
In this section, we always consider the simplest dressing factor, which contains only a pair of poles on the imaginary axis, i.e.
\begin{equation}\label{eq:definition_chi_general_review_2}
\chi_k(\lambda)=I+i\frac{\mu_k+\mu_k^{-1}}{\lambda-i\mu_k}g_{k-1}\frac{J W_k W_k^T J}{W_k^T g^{-1}_{k-1}W_k}-i\frac{\mu_k+\mu_k^{-1}}{\lambda+i\mu_k^{-1}}\frac{W_k W_k^T}{W_k^Tg^{-1}_{k-1}W_k}g_{k-1}^{-1},
\end{equation}
where
\begin{equation} \label{eq:Wk_definition}
W_k=\Psi_k(i\mu_k)p_k.
\end{equation}
This expression generalizes the dressing factor \eqref{eq:definition_chi_simplest_W}. The subscript $k$ is used as index for the location of the poles, as well as the corresponding constant vector $p,$ which appear in the dressing factor $\chi_k$. We remind the reader that these constant vectors should be real and null, i.e. $p^T_k J p_k=0$.  The element of the coset that corresponds to the new NLSM solution is
\begin{equation}
g_k=\Psi_{k+1}(0)=\chi_{k}(0)g_{k-1}.
\end{equation}
Putting everything together, the new element of the coset is
\begin{equation}
g_k=g_{k-1}-\frac{\mu_k+\mu_k^{-1}}{\mu_k}g_{k-1}\frac{J W_k W_k^T J}{W_k^Tg^{-1}_{k-1}W_k}g_{k-1}-\frac{\mu_k+\mu_k^{-1}}{\mu_k^{-1}}\frac{W_k W_k^T}{W_k^Tg^{-1}_{k-1}W_k}.
\end{equation}
This new element of the coset corresponds to a vector in the enhanced space of H$^3$ through the relation
\begin{equation}\label{eq:gk_mapping}
g_k=J+2J Y_k Y^T_kJ .
\end{equation}
Using this mapping, combined with the fact that $W_k^TJW_k=0$, it is trivial to show that
\begin{equation}
g_k=J-2J\left(\frac{Y_{k-1}}{\mu_k}+\frac{\mu_k+\mu_k^{-1}}{2}\frac{JW_k}{W_k^TY_{k-1}}\right)\left(\frac{Y_{k-1}}{\mu_k}+\frac{\mu_k+\mu_k^{-1}}{2}\frac{JW_k}{W_k^TY_{k-1}}\right)^TJ.
\end{equation}
Finally, in view of \eqref{eq:gk_mapping}, the new solution of the NLSM is
\begin{equation}\label{eq:Y_k}
Y_{k}=i\left(\frac{Y_{k-1}}{\mu_k}+\frac{\mu_k+\mu_k^{-1}}{2}\frac{JW_k}{W_k^TY_{k-1}}\right),
\end{equation}
where $W_k=\Psi_k(i\mu_k)p_k$. It is evident that successive dressing transformations indeed lead to an interchange of real and imaginary solutions of the NLSM.

The imaginary vector $Y_{k},$ normalized as $Y_k^TJY_k=-1$ is a solution of the equations of motion
\begin{equation}
\partial_+\partial_-Y_k-\left(\partial_+Y_k^TJ\partial_-Y_k\right)Y_k=0 ,
\end{equation}
which in addition satisfies the Virasoro constraints
\begin{equation}
\partial_\pm Y_k^TJ\partial_\pm Y_k=0.
\end{equation} 
Its imaginary part $\tilde{Y}_{k}$ is normalized as $\tilde{Y}_{k}^TJ\tilde{Y}_{k}=1,$ solves the equations of motion
\begin{equation}
\partial_+\partial_-\tilde{Y}_k+\left(\partial_+\tilde{Y}_k^TJ\partial_-\tilde{Y}_k\right)\tilde{Y}_k=0
\end{equation}
and it satisfies the Virasoro constraints
\begin{equation}
\partial_\pm \tilde{Y}_k^TJ\partial_\pm \tilde{Y}_k=0.
\end{equation}
Clearly, the imaginary part of the solution is a bona fide real solution of the NLSM defined on de Sitter space. The above analysis does not rely on the dimensionality of the enhanced space. \emph{Thus, a single dressing transformation with the simplest dressing factor in the coset $SO(1,d)/SO(d)$ interrelates solutions of the Euclidean NLSM on Hyperbolic space $\mathrm{H}^d$ and of the Euclidean NLSM on de Sitter space $\mathrm{dS}_d$}. This calculation reveals that in the case of Euclidean world-sheet coordinates, the dressing method may interrelate real solutions of different equations in general. This is analogous to B\"acklund transformations that connect solutions of different equations. 

By decomposing to the temporal and spatial components of the vectors $Y_{k}$ and $W_k$, we obtain
\begin{align}
Y_{k}^0&=i\left(\frac{Y^0_{k-1}}{\mu_k}-\frac{\mu_k+\mu_k^{-1}}{2}\frac{1}{Y^0_{k-1}+\vec{n}_k\cdot\vec{Y}_{k-1}}\right),\label{eq:y_k_temp}\\
\vec{Y}_{k}&=i\left(\frac{\vec{Y}_{k-1}}{\mu_k}+\frac{\mu_k+\mu_k^{-1}}{2}\frac{\vec{n}_k}{Y^0_{k-1}+\vec{n}_k\cdot\vec{Y}_{k-1}}\right)\label{eq:y_k_spat},
\end{align}
where
\begin{equation}
\vec{n}_k=\frac{\vec{W}_k}{W^0_k}.
\end{equation}
is a unit norm 3-vector. It is worth noticing that the solutions depend only on this vector and $Y_{k-1}$. Using equations \eqref{eq:y_k_temp} and \eqref{eq:y_k_spat}, along with \eqref{eq:psi_k} and \eqref{eq:definition_chi_general_review_2}, one can construct iteratively a whole tower of solutions without solving any equation or imposing any constraint.

It can be shown that the dressed solution obeys the equations of motion, as well as the Virassoro constraints, see appendix \ref{sec:appendix}. 

\subsection{The Tower of Real Solutions}

As already discussed, an even number of dressing transformations is needed, in order to obtain real solutions of the NLSM out of a real seed solution. Using \eqref{eq:Y_k} twice it is straightforward to show that the vector $Y_k$ reads
\begin{multline}\label{eq:Yk_2}
Y_k=\left(1-\frac{1+\mu_{k-1}^{-1}\mu_{k}^{-1}}{X}\right)Y_{k-2}\\+\frac{1}{2X}\frac{1+\mu_{k-1}\mu_{k}}{\mu_{k}-\mu_{k-1}}\left[\left(\mu_k+\mu_k^{-1}\right)\frac{JV_k}{V_k^T Y_{k-2}}-\left(\mu_{k-1}+\mu_{k-1}^{-1}\right)\frac{JV_{k-1}}{V_{k-1}^T Y_{k-2}}\right],
\end{multline}
where
\begin{equation}\label{eq:def_X}
X=1+\frac{1}{2}\frac{\left(1+\mu_k^2\right)\left(1+\mu_{k-1}^2\right)}{\left(\mu_k-\mu_{k-1}\right)^2}\frac{V_{k}^TJV_{k-1}}{\left(V_k^TY_{k-2}\right)\left(V_{k-1}^TY_{k-2}\right)}
\end{equation}
and
\begin{equation}\label{eq:def_Vk}
V_k=\Psi_{k-1}(i\mu_{k})p_{k},\quad V_{k-1}=\Psi_{k-1}(i\mu_{k-1})p_{k-1}.
\end{equation}
The null vectors $V$ are expressed in terms of $\Psi_{k-1}$ solely\footnote{The indices of the vectors $V$ are associated to the indices of the poles and the constant vectors.}. They should not be confused with the vectors $W$, but they are related to them via
\begin{equation}
W_{k-1}=V_{k-1}, \quad W_k = \chi_{k-1} \left( i \mu_k \right) V_k .
\end{equation}
The equation \eqref{eq:Yk_2} is symmetric under the transformation $\left(\mu_{k-1},p_{k-1}\right)\leftrightarrow\left(\mu_{k},p_{k}\right)$ in accordance with the expected permutability of the dressing transformations.

\section{Properties of the Dressed Static Minimal Surfaces}
\label{sec:properties}

In this section, we study some basic properties of the dressed minimal surfaces. For this purpose, we follow the approach introduced in \cite{Katsinis:2018ewd}, expressing the vector $Y$ as a matrix acting on a constant vector. Furthermore, in order to facilitate the solution of the auxiliary system for the specific example of the elliptic solutions, it is advantageous to write the equations of the auxiliary system in terms of the real coordinates $u$ and $v$, instead of the complex coordinates $z$ and $\bar z$.

The auxiliary system assumes the form
\begin{equation}
\partial_i\Psi(\lambda)=\left(\tilde{\partial}_i g\right)g^{-1}\Psi(\lambda),
\end{equation}
where $i=u,v$ and
\begin{equation}
\tilde{\partial}_u=\frac{1}{1-\lambda^2}\partial_u+i\frac{\lambda}{1-\lambda^2}\partial_v,\qquad\tilde{\partial}_v=-i\frac{\lambda}{1-\lambda^2}\partial_u+\frac{1}{1-\lambda^2}\partial_v.
\end{equation}
We express the seed solution $Y$ as a matrix $U(u,v)$ acting on a constant vector $\hat{Y},$ i.e. 
\begin{equation}
Y:=U\hat{Y}.
\label{eq:aux_U_def}
\end{equation}
The seed solution can be expressed as
\begin{equation}\label{eq:ghat_definition}
g= \theta U\theta \hat{g}JU^TJ,
\end{equation}
where the matrix $U$ must obey the property $U^{-1}=J U^{T}J$ so that
\begin{equation}\label{eq:ghat_mapping}
\hat{g}=\theta\left(I+2\hat{Y}\hat{Y}^TJ\right)
\end{equation}
is an element of the coset SO$(1,3)/$SO$(3)$. This also implies that $\hat{Y}$ belongs in H$^3$. In a similar manner, we define $\hat{\Psi}$ as
\begin{equation}\label{eq:psihat_definition}
\Psi:=\theta U \theta\hat{\Psi}.
\end{equation}
The auxiliary system assumes the form 
\begin{equation}
\partial_i\hat{\Psi}=\left\{\theta J U^T J\left[\left(\tilde{\partial}_i-\partial_i\right)U\right]\theta-\hat{g}J U^T J\left[\tilde{\partial}_iU\right]\hat{g}^{-1}+\left[\tilde{\partial}_i\hat{g}\right]\hat{g}^{-1}\right\}\hat{\Psi}
\end{equation}
in terms of the hatted quantities. Notice that, as $J U^T J =U^{-1}$, the form of the equations is identical to the ones derived in \cite{Katsinis:2018ewd}. As we already discussed, the choice \eqref{eq:special_representative} for $Y_0$ implies that $\theta=J$. In addition, one can select the matrix $U$ so that $\hat{Y}=Y_0$.  These choices set $\hat{g}=I$. Then, the equation of the auxiliary system simplifies to
\begin{equation}\label{eq:auxiliary_psi_hat}
\partial_i\hat{\Psi}=\left\{U^T J\left[\left(\tilde{\partial}_i-\partial_i\right)U\right]J-J U^T J\left[\tilde{\partial}_iU\right]\right\}\hat{\Psi},
\end{equation}
while the condition $\Psi(0)=g,$ reduces to
\begin{equation}\label{eq:Psi_initial_condition}
\hat{\Psi}(0)=J U^T J.
\end{equation}

\subsection{Geometric Depiction of the Dressing}

Expressing the solution \eqref{eq:Y_k} in terms of hatted quantities yields
\begin{equation}\label{eq:Y_hat_k}
\hat{Y}_{k}=i\left(\frac{\hat{Y}_{k-1}}{\mu_k}+\frac{\mu_k+\mu_k^{-1}}{2}\frac{J\hat{W}_k}{\hat{W}_k^T\hat{Y}_{k-1}}\right),\qquad \hat{W}_k=\hat{\Psi}_k\left(i \mu_k\right)p_k.
\end{equation}
In order to shed some light on the effect of the dressing transformation on the seed solution, we consider a single dressing transformation. For $k=1$, decomposing this vector to it's temporal and spatial components yields
\begin{align}
\hat{Y}_{1}^0&=-i\frac{\mu_1-\mu_1^{-1}}{2}\\
\vec{\hat{Y}}_{1}&=i\frac{\mu_1+\mu_1^{-1}}{2}\hat n_1,
\end{align}
where $\hat n_1=\vec{\hat W}_1/\hat W^0_1$ is a unit vector. Without loss of generality, we assume that $\mu_1$ is positive and we identify the quantity 
\begin{equation}
\vec\zeta \left( u , v \right) =-\left(\ln \mu_1-i\frac{\pi}{2}\right)\hat n_1 \left( u , v \right) ,
\end{equation}
as the rapidity of the Lorentz transformation
\begin{equation}
\Lambda(\vec\zeta)=\begin{pmatrix}
\cosh\zeta&\sinh\zeta\hat n_1^T\\
\sinh\zeta\hat n_1&I+\left(\cosh\zeta-1\right)\hat n_1\hat n_1^T
\end{pmatrix},
\end{equation}
which relates $\hat Y_0$ with $\hat Y_1$.

The physical reason for the interrelation between solutions of the NLSM in H$^d$ and solutions of the NLSM in dS$_d$ is the fact that  the particular dressing factor \eqref{eq:definition_chi_simplest_W} acts as a boost on $Y_0$ along the direction $-\hat{n}_1$ with \emph{superluminal} velocity of \emph{constant magnitude} equal to
\begin{equation}
v_{\textrm{boost}}=\tanh\zeta=\coth\left(\ln\mu_1\right).
\end{equation}
This also implies that the dressed solution $Y_1$ is connected to the seed solution $Y_0$ via a Lorentz transformation, which depends on the world-sheet coordinates, however its \emph{trace is constant}. The hatted ``frame'' is a frame, where this Lorentz transformation can be expressed as a boost solely, and, thus, its constant trace can be identified as $2 \left( 1 + \cosh \zeta\right)$, where $\zeta$ is the rapidity of the boost.

The fact that the magnitude of the boost velocity does not depend on the world-sheet coordinates is the analogue of a similar property that appears in dressed classical string solutions on $\mathbb{R} \times \mathrm{S}^2$ \cite{Katsinis:2018ewd}. In this case the dressed solution is connected to its seed via a rotation, whose direction depends on the world-sheet coordinates, nevertheless the angle of the rotation is constant.

\subsection{On the Entangling Curve of the Dressed Minimal Surface}

The most basic property of the dressed minimal surface in the context of entanglement, is the form of the corresponding entangling surface and the relation of the latter with the one of the seed. In order to specify the entangling surface that corresponds to the dressed minimal surface, one needs to specify where the dressed solution $Y_{k}$ \eqref{eq:Yk_2} diverges. According to \eqref{eq:Yk_2}, a naive guess is that $Y_{k}$ may diverge due to a divergence of $Y_{k-2}$. The specific example of the dressed elliptic minimal surfaces, which is presented in section \ref{subsec:Doubly_Dressed_Elliptic}, indicates that the divergences of $Y_{k-2}$ are not inherited to $Y_{k}$. It is unclear whether this is always the case. This behavior is similar to the action of the dressing transformation on the elliptic strings. The dressed strings have spikes, as their precursors, but the spikes  do not appear at the same locations as in the seeds\cite{Katsinis:2019oox}.

A divergence of $Y_k$ may emerge where $X$ vanishes. Since $Y_{k-2}$ is timelike, one can always select a matrix $\mathcal{U}\in\textrm{SO}(1,3)$, so that $Y_{k-2}=\mathcal{U}Y_0$, where $Y_0$ is given by \eqref{eq:special_representative}. Similarly we define 
\begin{equation}\label{eq:def_V_tilde}
\tilde{V}_k=J\mathcal{U}^TJ V_k.
\end{equation}
Then, equation \eqref{eq:def_X} assumes the form
\begin{equation}\label{eq:def_X_nhat}
X=1+\frac{1}{2}\frac{\left(1+\mu_k^2\right)\left(1+\mu_{k-1}^2\right)}{\left(\mu_k-\mu_{k-1}\right)^2}\left(-1+\hat{n}_k\cdot \hat{n}_{k-1}\right),
\end{equation}
where
\begin{equation}\label{eq:nhat}
\hat{n}_k=\frac{\vec{\tilde{V}}_k}{\tilde{V}^0_k},\qquad \hat{n}_{k-1}=\frac{\vec{\tilde{V}}_{k-1}}{\tilde{V}^0_{k-1}}
\end{equation}
are unit vectors since $V_k$ and $V_{k-1}$ are null. Furthermore, because
\begin{equation}
\frac{\left(1+\mu_k^2\right)\left(1+\mu_{k-1}^2\right)}{\left(\mu_k-\mu_{k-1}\right)^2}\geq 1,
\end{equation}
it is possible for $X$ to vanish, thus (at least part of) the boundary region may be specified by the equation $X=0$.

Finally, $Y_{k}$ could diverge when the term $V_k/\left(V_k^T Y_{k-2}\right)$ or the similar term with $V_k\rightarrow V_{k-1}$ diverges. Since $V_k$ is null we obtain
\begin{equation}
\frac{V_k}{V_k^T Y_{k-2}}=\frac{1}{-Y_{k-2}^0+\frac{\vec{V}_k}{V^0_k}\cdot\vec{Y}_{k-2}}\begin{pmatrix}
1 \\
\frac{\vec{V}_k}{V^0_k}
\end{pmatrix},
\end{equation}
where $\vec{V}_k/V^0_k$ is a unit vector. As $Y_{k-2}$ is timelike $\vert Y^0_{k-2}\vert\geq \vert \vec{Y}_{k-2}\vert$, this term is regular unless $Y_{k-2}$ diverges. Thus, the boundary of the dressed minimal surface $Y_k$ is potentially obtained for the same subset of the world-sheet coordinates that correspond to the boundary of $Y_{k-2}$ or to the solutions of the equation $X=0$.

\subsection{The Surface Element of the Dressed Minimal Surface}
In view of the Ryu and Takayanagi prescription for the computation of the holographic entanglement entropy, the calculation of the area of the dressed minimal surface presents a certain interest. The surface element of the dressed minimal surface, which is provided by equation \eqref{eq:pohl_addition}, can be re-expressed through the use of the identity
\begin{equation}
\frac{\partial_+f\partial_-f}{f^2}=\frac{\partial_+\partial_-f}{f}-\partial_+\partial_-\ln f
\end{equation}
along with \eqref{eq:WY_der2}, in the form
\begin{equation}\label{eq:algebraic_addition_1}
\left(\partial_+ Y_k\right)^T J \left(\partial_- Y_k\right) = \left(\partial_+ Y_{k-1}\right)^T J \left(\partial_- Y_{k-1}\right) - \partial_+\partial_-\ln\left[\left(W_k^T Y_{k-1}\right)^2\right].
\end{equation}
The latter provides an algebraic addition formula that relates the surface element of the dressed minimal surface with the surface element of its seed. Since we are interested in a relation between real solutions of the NLSM, we can express this addition formula as 
\begin{multline}\label{eq:algebraic_addition_2}
\left(\partial_+ Y_k\right)^T J \partial_- Y_k=\left(\partial_+ Y_{k-2}\right)^T J \partial_- Y_{k-2}-\partial_+\partial_-\ln\left[\left(\left(V_k^TY_{k-2}\right)\left(V_{k-1}^TY_{k-2}\right)X\right)^2\right],
\end{multline}
where $X$ is given by \eqref{eq:def_X}. As already discussed, unless $Y_{k-2}$ diverges, $V_{k-1}^TY_{k-2}$ and $V_k^TY_{k-2}$ do not vanish, since these terms are the inner product of a null vector with as timelike one. Let us denote $\mathcal{D}_k$ the domain of the world-sheet coordinates of the dressed minimal surface $Y_k$. Assuming that the boundary of this surface corresponds only to the solutions of the equations $X=0$ and $\mathcal{D}_k$ does not contain divergences of $Y_{k-2}$. Then, the area of the dressed minimal surface is 
\begin{equation}
A_k=\int_{\mathcal{D}_k} du dv \left(\partial_+ Y_{k-2}\right)^T J \partial_- Y_{k-2}- \int_{\mathcal{D}_k}du dv\nabla^2\ln\left[\left(\left(V_k^TY_{k-2}\right)\left(V_{k-1}^TY_{k-2}\right)X\right)^2\right]
\end{equation}
or using Green's identity
\begin{equation}
A_k=\int_{\mathcal{D}_k} du dv \left(\partial_+ Y_{k-2}\right)^T J \partial_- Y_{k-2}- \int_{\partial\mathcal{D}_k}d\ell \hat{n}\cdot\vec{\nabla}\ln\left[\left(\left(V_k^TY_{k-2}\right)\left(V_{k-1}^TY_{k-2}\right)X\right)^2\right].
\end{equation}
\section{Dressed Static Elliptic Minimal Surfaces in AdS$_4$}
\label{sec:Dressed_Elliptic}
In this section, we apply the dressing method, considering the elliptic minimal surfaces \cite{Pastras:2016vqu} as the seed solution, in order to construct new static minimal surfaces in AdS$_4$.

\subsection{Elliptic Minimal Surfaces}

Very few minimal surfaces are known in a form that can be used for the computation of their area. This picture changes drastically in the case of static minimal surfaces in AdS$_4$, where the whole class of elliptic minimal surfaces has been constructed in \cite{Pastras:2016vqu}. Therein, the author exploits the fact that co-dimension two minimal surfaces in AdS$_4$ extremize a NLSM action, to relate the static minimal surfaces via Pohlmeyer reduction to solutions of the Euclidean $\cosh$-Gordon equation. In particular, the author considers the elliptic solutions of the $\cosh$-Gordon equation, which possess the property that they depend solely on one out of the two isothermal, world-sheet coordinates, which parametrize the surface. Subsequently, the Pohlmeyer mapping is inverted, which leads to the construction of the static elliptic minimal surfaces in a simple handy form. The aforementioned inversion is in general non-trivial due to the fact that Pohlmeyer reduction constitutes a many to one, non-local mapping. Moreover, it is shown that the Pohlmeyer field is related to the area of the minimal surface, which renders the computation of the area straightforward. 

The solutions of the Euclidean $\cosh$-Gordon equation that depend only on $u$ read
\begin{equation}
\alpha=\ln\left[2\left(\wp\left(u;g_2,g_3\right)-e_2\right)\right], 
\end{equation}
where $\wp\left(u;g_2,g_3\right)$ is the Weierstrass elliptic function with moduli $g_2$ and $g_3$. The moduli are expressed in terms of a real integration constant $E$ through the relations. 
\begin{equation}
g_2=\frac{E^2}{3}+1\quad \text{and}\quad g_3=-\frac{E}{3}\left(\frac{E^2}{9}+\frac{1}{2}\right).
\label{eq:elliptic_moduli}
\end{equation}
The roots of the associated cubic polynomial assume the form
\begin{equation}
e_1=-\frac{E}{12}+\sqrt{\left(\frac{E}{4}\right)^2+\frac{1}{4}}, \quad
e_2=\frac{E}{6}, \quad
e_3=-\frac{E}{12}-\sqrt{\left(\frac{E}{4}\right)^2+\frac{1}{4}}
\end{equation}
and they obey $e_1>e_2>e_3$.

The static minimal surfaces in AdS$_4$, that correspond to the above solutions of the Euclidean $\cosh$-Gordon equation, are parametrized as follows:
\begin{equation}
Y=\begin{pmatrix}
F_1(u)\cosh\left(\varphi_1(u,v)\right)\\
F_1(u)\sinh\left(\varphi_1(u,v)\right)\\
F_2(u)\cos\left(\varphi_2(u,v)\right)\\
F_2(u)\sin\left(\varphi_2(u,v)\right)
\end{pmatrix},
\label{eq:elliptic_minimal_surface}
\end{equation}
where
\begin{equation}
F_1(u)=\frac{\sqrt{\wp(u)-\wp(a_1)}}{\sqrt{\wp(a_2)-\wp(a_1)}}, \quad
F_2(u)=\frac{\sqrt{\wp(u)-\wp(a_2)}}{\sqrt{\wp(a_2)-\wp(a_1)}}
\label{eq:elliptic_solutios_Fs}
\end{equation}
and
\begin{equation}
\varphi_1(u,v)=\ell_1 v +\phi_1(u),\quad
\varphi_2(u,v)=\ell_2 v -\phi_2(u),\label{eq:elliptic_solution_varphis}
\end{equation}
where
\begin{align}
\ell_1=\sqrt{\wp(a_2)-e_2}, &\quad \phi_1(u)=\frac{1}{2}\ln\left(-\frac{\sigma(u+a_1)}{\sigma(u-a_1)}\right)-\zeta\left(a_1\right)u,\label{eq:phi_1}\\
\ell_2=\sqrt{e_2-\wp(a_1)}, &\quad \phi_2(u)=-\frac{i}{2}\ln\left(-\frac{\sigma(u+a_2)}{\sigma(u-a_2)}\right)+i\zeta\left(a_2\right)u\label{eq:phi_2}.
\end{align}
The functions $\zeta(u)$ and $\sigma(u)$ are the Weierstrass zeta and sigma functions. 

The parameters $\wp(a_1)$ and $\wp(a_2)$ are not both free, but they are subject to the constraint
\begin{equation}
\wp(a_1)+\wp(a_2)=-e_2,\label{eq:Virasoro_el_1}
\end{equation}
whereas their relative sign is determined by the equation
\begin{equation}
\wp^\prime(a_1) \ell_1+i \wp^\prime(a_2) \ell_2=0 .\label{eq:Virasoro_el_2}
\end{equation}
Their range obeys the inequalities
\begin{equation}
e_1>\wp(a_2)>e_2 , \quad
e_2>\wp(a_1)>e_3.
\label{eq:elliptic_range}
\end{equation}

The range of the coordinates, which corresponds to a single minimal surface with a connected boundary, $u$ and $v$ is 
\begin{equation}
u\in\left(2n\omega_1,2(n+1)\omega_1\right),\quad v\in\mathbb R,\quad\text{where}\quad n\in\mathbb{Z}
\end{equation}
and $\omega_1$ is the real half-period of the Weierstrass elliptic function, given the moduli \eqref{eq:elliptic_moduli}. The boundary region of the minimal surface \eqref{eq:elliptic_minimal_surface} lies at $u=2n\omega_1$, with $n\in\mathbb Z$, while the area of the minimal surface, which is of great interest for the computation of the holographic entanglement entropy, is given by the expression
\begin{equation} 
A=\int_{-\infty}^{+\infty}dv\int_{2n\omega_1}^{2(n+1)\omega_1}du\left(\wp(u)-e_2\right).
\end{equation}

Some interesting limits of the minimal surface \eqref{eq:elliptic_minimal_surface} are the helicoid, the catenoid and the cusp limit. The helicoid minimal surface is obtained when the quantities $\wp(\alpha_1)$ and $\wp(\alpha_2)$ assume the values $e_3$ and $e_1$ respectively, independently of the sign of $E$. When $\wp(\alpha_1)=e_2$ and $E>0$ the minimal surface reduces to the catenoid. Finally, the cusp limit corresponds to $\wp(\alpha_1)=e_2$ and $E<0$. For further details on the construction of the static elliptic minimal surfaces in AdS$_4$, the reader is referred to \cite{Pastras:2016vqu}.

\subsection{The Auxiliary System}

The elliptic minimal surfaces have a particular dependence on the real world-sheet coordinates $u$ and $v$. More specifically, the dependence on the coordinate $v$ is very simple, due to the fact that the Pohlmeyer counterpart does not depend on $v$ at all. Therefore, it is advantageous to express the auxiliary system in terms of the real coordinates $u$ and $v$ in the form \eqref{eq:auxiliary_psi_hat}, instead of the original formulation in terms of the complex coordinates $z$ and $\bar z$ \eqref{eq:aux_formal}.

The form of the static elliptic minimal surfaces \eqref{eq:elliptic_minimal_surface} implies that the matrix $U$, which connects $Y$ to $\hat{Y}$, through the equation \eqref{eq:aux_U_def} can be written as
\begin{equation}\label{eq:rotation}
U=U_2U_1,
\end{equation}
where
\begin{align}
U_1&=\begin{pmatrix}
F_1 & 0 & F_2 & 0\\
0   & 1 & 0   & 0\\
F_2 & 0 & F_1 & 0\\
0   & 0 & 0   & 1\\
\end{pmatrix},\\
U_2&=\begin{pmatrix}
\cosh\left(\varphi_1(u,v)\right) & \sinh\left(\varphi_1(u,v)\right) & 0 & 0\\
\sinh\left(\varphi_1(u,v)\right)   & \cosh\left(\varphi_1(u,v)\right) & 0   & 0\\
0 & 0 & \cos\left(\varphi_2(u,v)\right) & -\sin\left(\varphi_2(u,v)\right)\\
0   & 0 & \sin\left(\varphi_2(u,v)\right)   & \cos\left(\varphi_2(u,v)\right)\\
\end{pmatrix} .
\end{align}

In order to proceed we must obtain specific expressions for the derivatives that appear in \eqref{eq:auxiliary_psi_hat} using the explicit form of the static elliptic minimal surfaces \eqref{eq:elliptic_minimal_surface}. Following equations \eqref{eq:elliptic_solutios_Fs} and \eqref{eq:elliptic_solution_varphis}, the derivatives of the various functions that appear in \eqref{eq:elliptic_minimal_surface} obey the following relations
\begin{equation}
\partial_v F_i =0,\quad \partial_u F_i =\frac{F_3}{F_i},\quad \textrm{where} \quad F_3=\frac{\wp^\prime(u)}{2\left(\wp(a_2)-\wp(a_1)\right)}
\label{eq:aux_F_derivatives}
\end{equation}
and
\begin{equation}
\partial_v \varphi_i =\ell_i, \quad \partial_u \varphi_1 =\phi^\prime_1 = -\frac{1}{2}\frac{\wp^\prime(a_1)}{\wp(u)-\wp(a_1)}, \quad
\partial_u \varphi_2 = -\phi^\prime_2 = - \frac{i}{2}\frac{\wp^\prime(a_2)}{\wp(u)-\wp(a_2)}.
\label{eq:aux_varphi_derivatives}
\end{equation}

We introduce the generators of the $\mathrm{SO}(1,3)$ group
\begin{align}
K_1&=\begin{pmatrix}
0 & 1 & 0 & 0\\
1 & 0 & 0 & 0\\
0 & 0 & 0 & 0\\
0 & 0 & 0 & 0
\end{pmatrix},\quad \phantom{-}K_2=\begin{pmatrix}
0 & 0 & 1 & 0\\
0 & 0 & 0 & 0\\
1 & 0 & 0 & 0\\
0 & 0 & 0 & 0
\end{pmatrix},\quad \phantom{-}K_3=\begin{pmatrix}
0 & 0 & 0 & 1\\
0 & 0 & 0 & 0\\
0 & 0 & 0 & 0\\
1 & 0 & 0 & 0
\end{pmatrix},\\ T_1&=\begin{pmatrix}
0 & 0 & 0 & 0\\
0 & 0 & 0 & 0\\
0 & 0 & 0 & -1\\
0 & 0 & 1 & 0
\end{pmatrix},\quad
T_2=\begin{pmatrix}
0 & 0 & 0 & 0\\
0 & 0 & 0 & 1\\
0 & 0 & 0 & 0\\
0 & -1 & 0 & 0
\end{pmatrix},\quad
T_3=\begin{pmatrix}
0 & 0 & 0 & 0\\
0 & 0 & -1 & 0\\
0 & 1 & 0 & 0\\
0 & 0 & 0 & 0
\end{pmatrix},\label{eq:generators_KT}
\end{align}
in order to express the auxiliary system in the form
\begin{equation}
\partial_i\hat{\Psi}=\left(\kappa_i^jK_j+\tau_i^jT_j\right)\hat{\Psi}.
\end{equation}
Using the equations \eqref{eq:aux_F_derivatives} and \eqref{eq:aux_varphi_derivatives}, it is a matter of algebra to show that
\begin{align}
\kappa_u^1&=-F_1\left(\frac{1+\lambda^2}{1-\lambda^2}\phi_1^\prime+i \frac{2\lambda}{1-\lambda^2}\ell_1\right), \label{eq:value_ku1}\\
\kappa_u^2&=-\frac{1+\lambda^2}{1-\lambda^2}\frac{F_3}{F_1F_2},\\
\kappa_u^3&=-F_2\left(-\frac{1+\lambda^2}{1-\lambda^2}\phi_2^\prime+i \frac{2\lambda}{1-\lambda^2}\ell_2\right)
\end{align}
and
\begin{equation}
\tau_u^1=F_1\phi_2^\prime,\quad
\tau_u^2=0,\quad
\tau_u^3=F_2\phi_1^\prime ,
\end{equation}
as well as,
\begin{align}
\kappa_v^1&=-F_1\left(\frac{1+\lambda^2}{1-\lambda^2}\ell_1-i \frac{2\lambda}{1-\lambda^2}\phi_1^\prime\right),\\
\kappa_v^2&=i\frac{2\lambda}{1-\lambda^2}\frac{F_3}{F_1F_2},\\
\kappa_v^3&=-F_2\left(\frac{1+\lambda^2}{1-\lambda^2}\ell_2+i \frac{2\lambda}{1-\lambda^2}\phi_2^\prime\right)
\end{align}
and
\begin{equation}
\tau_v^1=-\ell_2F_1,\quad
\tau_v^2=0,\quad
\tau_v^3=\ell_1F_2\label{eq:value_tv}.
\end{equation}
The vectors $\vec{\kappa}_u$, $\vec{\kappa}_v$, $\vec{\tau}_u$ and $\vec{\tau}_v$ do not depend on the coordinate $v$. Under the inversion of $\lambda$ these quantities have the following parity properties
\begin{equation}
\kappa_i^j(1/\lambda)=-\kappa_i^j(\lambda),\qquad \tau_i^j(1/\lambda)=\tau_i^j(\lambda).
\end{equation}
Under complex conjugation, they also obey 
\begin{equation}
\bar{\kappa}_j^i(\bar{\lambda})=\kappa_j^i(-\lambda),\qquad \bar{\tau}_i^j(\bar{\lambda})=\tau_i^j(-\lambda).
\end{equation}

The vectors $\vec{\kappa}_u$, $\vec{\kappa}_v$, $\vec{\tau}_u$ and $\vec{\tau}_v$ obey a set of properties that will be handy in what follows. The first one is the fact that the inner product $\vec{\kappa}_v\cdot\vec{\tau}_v := \delta_1$ does not depend on the world-sheet coordinates. Using the equations \eqref{eq:elliptic_solutios_Fs}, \eqref{eq:aux_varphi_derivatives} and equations \eqref{eq:value_ku1} to \eqref{eq:value_tv}, as well as the property $F_1^2-F_2^2=1$, it is straightforward to calculate that
\begin{equation}\label{eq:delta1_value}
\delta_1=\frac{1+\lambda^2}{1-\lambda^2}\ell_1\ell_2+\frac{2i\lambda}{1-\lambda^2}\frac{\wp^\prime(a_1)}{2\ell_2}.
\end{equation}
Similarly, the quantity $\vert \vec{\kappa}_v\vert^2-\vert \vec{\tau}_v\vert^2 := \delta_2$ is also constant. It is a matter of tedious algebra to show that
\begin{equation}\label{eq:delta2_value}
\delta_2=-3e_2.
\end{equation}
In a similar manner, the inner product $\vec{\kappa}_v\cdot\vec{\tau}_v := \delta_3$ does not depend on the world-sheet coordinates,
\begin{equation}\label{eq:delta3_value}
\delta_3=\frac{2i\lambda}{1-\lambda^2}\ell_1\ell_2-\frac{1+\lambda^2}{1-\lambda^2}\frac{\wp^\prime(a_1)}{2\ell_2}.
\end{equation}
Finally, the vectors $\vec{\kappa}_u$ and $\vec{\kappa}_v$ obey,
\begin{equation}\label{eq:delta4_value}
\vec{\kappa}_u\cdot\vec{\kappa}_v=0.
\end{equation}
The fact that $\vec{\kappa}_v$ and $\vec{\kappa}_u$ are perpendicular is not accidental: it can be shown that the above inner product vanishes as a direct consequence of the Virasoro constraints. The constants $\delta_1$ and $\delta_3$ also satisfy
\begin{equation}
\delta_1^2+\delta_3^2=(e_1-e_2)(e_2-e_3)=\frac{1}{4}.\label{eq:delta_13_sq}
\end{equation}
This relation will become important in what follows.

\subsection{The Solution of the Auxiliary System}
\label{subsec:auxiliary_solution_elliptic}
The auxiliary system \eqref{eq:auxiliary_psi_hat} for the matrix $\hat{\Psi}$ can be decomposed into four independent, identical equations for its columns $\hat{\Psi}_i$. Since $\kappa_v^i$ and $\tau_v^i$ do not depend on the variable $v$, one can solve the set of equations
\begin{equation}\label{eq:equations_v}
\partial_v\hat{\Psi}_i=\left(\kappa_v^jK_j+\tau_v^jT_j\right)\hat{\Psi}_i,
\end{equation}
as a system of ordinary differential equations with constant coefficients and promote the integration constants to arbitrary functions of the variable $u$. These functions will be specified using the remaining equations of the auxiliary system, i.e. those that involve $\kappa_u^i$ and $\tau_u^i$.

The matrix on the right-hand-side of \eqref{eq:equations_v}, i.e. $\left(\kappa_v^jK_j+\tau_v^jT_j\right)$, has four distinct eigenvalues, namely the solutions of the equation
\begin{equation}\label{eq:eigenvalues_eq}
\Lambda^4-\Lambda^2\delta_2-\delta_1^2=0.
\end{equation}
We will denote these eigenvalues as $\Lambda_{\pm 1}$ and $\Lambda_{\pm 2}$. They are equal to
\begin{align}
\Lambda_{\pm 1} = \pm L_1 , &\quad \textrm{where} \quad L_1(\lambda)=\frac{1}{\sqrt{2}}\sqrt{\sqrt{4\delta_1^2+\delta_2^2}+\delta_2},\\
\Lambda_{\pm 2} = \pm i L_2 , &\quad \textrm{where} \quad L_2(\lambda)=\frac{1}{\sqrt{2}}\sqrt{\sqrt{4\delta_1^2+\delta_2^2}-\delta_2}.
\end{align}
The quantities $\delta_1$ and $\delta_2$ are given by \eqref{eq:delta1_value} and \eqref{eq:delta2_value} respectively. We should mention that
\begin{equation}
\left.\left(4\delta_1^2+\delta_2^2\right)\right|_{\lambda=0}=\left(\wp(a_2)+2e_2\right)^2,
\end{equation}
and furthermore the quantity $\wp(a_2)+2e_2$ is always positive\footnote{For positive $E$ the range of $\wp(a_2)$ is $e_1>\wp(a_2)>e_2,$ while for negative $E$ the range is $e_1>\wp(a_2)>-2e_2$. Thus, in any case $2\wp(a_2)+e_2>3\vert e_2\vert$.}. Since
\begin{equation}
L_i(0)=\ell_i,
\end{equation}
these quantities are a natural generalization of the parameters $\ell_1$ and $\ell_2$ for the dressed solutions. In addition, under the inversion $\lambda\to\lambda^{-1}$, the eigenvalues obey
\begin{equation}
L_i(1/\lambda)=L_i(\lambda).
\end{equation}

The solution of the system of equations \eqref{eq:equations_v} assumes the form
\begin{equation}
\hat{\Psi}_i= \sum_k {C_i^k (u) V_k e^{\Lambda_k v}} ,
\label{eq:aux_solution_v}
\end{equation}
where $k$ takes the values $\pm 1$ and $\pm 2$. The vector $V_k$ is the eigenvector of the matrix $\left(\kappa_v^j K_j+\tau_v^jT_j\right)$ corresponding to the eigenvalue $\Lambda_k$; it is given by
\begin{equation}
V_k=\begin{pmatrix}
\Lambda_k\left(\vert \vec{\tau}_v\vert^2+\Lambda_k^2\right)\\
\tau_v^1\delta_1+\left(\tau_v^2\kappa_v^3-\tau_v^3\kappa_v^2\right)\Lambda_k+\kappa_v^1\Lambda_k^2 \\
\tau_v^2\delta_1+\left(\tau_v^3\kappa_v^1-\tau_v^1\kappa_v^3\right)\Lambda_k+\kappa_v^2\Lambda_k^2 \\
\tau_v^3\delta_1+\left(\tau_v^1\kappa_v^2-\tau_v^2\kappa_v^1\right)\Lambda_k+\kappa_v^3\Lambda_k^2
\end{pmatrix} .
\end{equation}
It will be convenient to express the spatial components of $V_k$ as
\begin{equation}
\vec{V}_k=\vec{\tau}_v\delta_1+\vec{\tau}_v\times\vec{\kappa}_v\Lambda_k+\vec{\kappa}_v\Lambda_k^2,
\end{equation}
in order to keep a more compact notation. Doing so, the eigenvectors read
\begin{equation}
V_k \equiv \begin{pmatrix} V_k^0\\
\vec{V}_k\end{pmatrix}=\begin{pmatrix}
\Lambda_k\left(\vert \vec{\tau}_v\vert^2+\Lambda_k^2\right)\\
\delta_1 \vec{\tau}_v + \Lambda_k \vec{\tau}_v\times\vec{\kappa}_v + \Lambda_k^2 \vec{\kappa}_v
\end{pmatrix}.
\end{equation}
With the aid of \eqref{eq:eigenvalues_eq}, it is easy to verify that the eigenvectors obey the properties,
\begin{equation}
V_{\pm i}^T J V_{\pm j}=0 , \quad V_{\pm i}^T J V_{\mp j} \propto \delta_{ij} .
\end{equation}
These relations imply that the four eigenvectors $V_k$ are linearly independent.

Substituting the equation \eqref{eq:aux_solution_v} into the yet unsolved equations of the auxiliary system 
\begin{equation}\label{eq:equations_u}
\partial_u\hat{\Psi}_i=\left(\kappa_u^jK_j+\tau_u^jT_j\right)\hat{\Psi}_i
\end{equation}
yields
\begin{equation}
\partial_u\left[C_i^k \begin{pmatrix} V_k^0\\
\vec{V}_k\end{pmatrix}\right]=\left(\kappa_u^jK_j+\tau_u^jT_j\right)\left[C_i^k\begin{pmatrix} V_k^0\\
\vec{V}_k\end{pmatrix}\right],
\end{equation}
since the four eigenvectors are linearly independent. In the following, we omit the subscripts $k$ and $i$ on $V^0$, $\vec{V}$, $C$ and $\Lambda$ for simplicity.
It is straightforward that this system of equations is equivalent to
\begin{equation}\label{eq:equations_u_vector}
\left[\partial_u\ln C \right]\begin{pmatrix} V^0\\
\vec{V}\end{pmatrix}+\begin{pmatrix} \partial_u V^0- \vec{\kappa}_u\cdot\vec{V}\\
\left(\partial_u-\vec{\tau}_u\times\right)\vec{V}-\vec{\kappa}_u V^0\end{pmatrix}=0.
\end{equation}

In order to solve the above, the derivatives of the coefficients $\kappa_v^i$ and $\tau_v^i$ with respect to the coordinate $u$, are required. It can be shown that they obey the following relations
\begin{align}
\partial_u \vec{\kappa}_v &= \vec{\kappa}_u\times\vec{\tau}_v-\vec{\kappa}_v\times\vec{\tau}_u , \label{eq:derivative_kv}\\
\partial_u \vec{\tau}_v &= \vec{\kappa}_v\times\vec{\kappa}_u-\vec{\tau}_v\times\vec{\tau}_u , \label{eq:derivative_tv}\\
\partial_u \vec{\kappa}_u &= - \vec{\kappa}_u\times\vec{\tau}_u - \vec{\kappa}_v\times\vec{\tau}_v .
\end{align}
These relations demonstrate why the quantities $\delta_i$ are constants, as well as the fact that the vectors $\vec{\kappa}_u$ and $\vec{\kappa}_v$ are perpendicular.

The temporal component of equation \eqref{eq:equations_u_vector} assumes the form
\begin{equation}
\partial_u\ln C = -\frac{2\vec{\tau}_v\cdot\partial_u\vec{\tau}_v}{\vert \vec{\tau}_v\vert^2+\Lambda^2}+\frac{\frac{\delta_1}{\Lambda} \vec{\kappa}_u\cdot\vec{\tau}_v +\vec{\kappa}_u\cdot\left(\vec{\tau}_v\times\vec{\kappa}_v\right)}{\vert \vec{\tau}_v\vert^2+\Lambda^2}.
\end{equation}
Taking into account \eqref{eq:derivative_kv}, which implies that $\vec{\kappa}_u\cdot\left(\vec{\tau}_v\times\vec{\kappa}_v\right)=\vec{\tau}_v\cdot\partial_u\vec{\tau}_v,$ along with \eqref{eq:delta3_value} we obtain that
\begin{equation}\label{eq:temporal_eq}
\partial_u\ln C =-\frac{\vec{\tau}_v\cdot\partial_u\vec{\tau}_v}{\vert \vec{\tau}_v\vert^2+\Lambda^2} + \frac{\delta_1 \delta_3}{\Lambda} \frac{1}{\vert \vec{\tau}_v\vert^2+\Lambda^2}.
\end{equation}

Before solving this equation, we will show that the spatial components of equation \eqref{eq:equations_u_vector} are redundant. Equations \eqref{eq:derivative_kv} and \eqref{eq:derivative_tv} imply that
\begin{multline}
\partial_u\vec{V} = \delta_1 \left[\vec{\kappa}_v\times\vec{\kappa}_u-\vec{\tau}_v\times\vec{\tau}_u\right] + \Lambda^2 \left[\vec{\kappa}_u\times\vec{\tau}_v-\vec{\kappa}_v\times\vec{\tau}_u\right]\\
+ \Lambda \left[\left(\vec{\kappa}_v\times\vec{\kappa}_u\right)\times\vec{\kappa}_v-\left(\vec{\tau}_v\times\vec{\tau}_u\right)\times\vec{\kappa}_v+\vec{\tau}_v\times\left(\vec{\kappa}_u\times\vec{\tau}_v\right)-\vec{\tau}_v\times\left(\vec{\kappa}_v\times\vec{\tau}_u\right)\right].
\end{multline}
Using the Jacobi identity on the triple cross products involving $\vec{\tau}_u$, it is straightforward to obtain that
\begin{equation}
\left(\partial_u-\vec{\tau}_u\times\right)\vec{V} = \delta_1 \vec{\kappa}_v\times\vec{\kappa}_u + \Lambda \left[\left(\vec{\kappa}_v\times\vec{\kappa}_u\right)\times\vec{\kappa}_v+\vec{\tau}_v\times\left(\vec{\kappa}_u\times\vec{\tau}_v\right)\right] + \Lambda^2 \vec{\kappa}_u\times\vec{\tau}_v.
\end{equation}
Then, its a matter of algebra to show that
\begin{equation}\label{eq:derivative_cov_v}
\left(\partial_u-\vec{\tau}_u\times\right)\vec{V}-V^0\vec{\kappa}_u=\Lambda\left(\vert\vec{\kappa}_u\vert^2-\Lambda^2\right)\vec{\kappa}_u-\Lambda\delta_3\vec{\tau}_v+\delta_1\vec{\kappa}_v\times\vec{\kappa}_u+\Lambda^2\vec{\kappa}_u\times\vec{\tau}_v.
\end{equation}
We decompose the vectors $\vec{\kappa}_u$, $\vec{\kappa}_v\times\vec{\kappa}_u$ and $\vec{\kappa}_u\times\vec{\tau}_v$ into the basis formed out of the vectors $\vec{\kappa}_v$, $\vec{\tau}_v$ and $\vec{\tau}_v\times\vec{\kappa}_v$ as follows:
\begin{align}
\vec{\kappa}_u&=\frac{\delta_3}{\vert\vec{\tau}_v\vert^2\vert\vec{\kappa}_v\vert^2-\delta_1^2}\left(\vert\vec{\kappa}_v\vert^2\vec{\tau}_v-\delta_1\vec{\kappa}_v\right)+\frac{\vec{\tau}_v\cdot\left(\vec{\kappa}_v\times\vec{\kappa}_u\right)}{\vert\vec{\tau}_v\vert^2\vert\vec{\kappa}_v\vert^2-\delta_1^2}\vec{\tau}_v\times\vec{\kappa}_v,\label{eq:expansion_ku}\\
\vec{\kappa}_v\times\vec{\kappa}_u&=\frac{\vec{\tau}_v\cdot\left(\vec{\kappa}_v\times\vec{\kappa}_u\right)}{\vert\vec{\tau}_v\vert^2\vert\vec{\kappa}_v\vert^2-\delta_1^2}\left(\vert\vec{\kappa}_v\vert^2\vec{\tau}_v-\delta_1\vec{\kappa}_v\right)-\frac{\delta_3\vert\vec{\kappa}_v\vert^2}{\vert\vec{\tau}_v\vert^2\vert\vec{\kappa}_v\vert^2-\delta_1^2}\vec{\tau}_v\times\vec{\kappa}_v,\label{eq:expansion_kvku}\\
\vec{\kappa}_u\times\vec{\tau}_v&=\frac{\vec{\tau}_v\cdot\left(\vec{\kappa}_v\times\vec{\kappa}_u\right)}{\vert\vec{\tau}_v\vert^2\vert\vec{\kappa}_v\vert^2-\delta_1^2}\left(\vert\vec{\tau}_v\vert^2\vec{\kappa}_v-\delta_1\vec{\tau}_v\right)+\frac{\delta_1\delta_3}{\vert\vec{\tau}_v\vert^2\vert\vec{\kappa}_v\vert^2-\delta_1^2}\vec{\tau}_v\times\vec{\kappa}_v.\label{eq:expansion_kutv}
\end{align}
By substituting equation \eqref{eq:temporal_eq}, as well as \eqref{eq:derivative_cov_v}, alongside with equations \eqref{eq:expansion_ku}, \eqref{eq:expansion_kvku} and \eqref{eq:expansion_kutv}, into the spatial component of equation \eqref{eq:equations_u_vector}, it is a matter of algebra to show that it is indeed satisfied.

We return to the solution of equation \eqref{eq:temporal_eq}. Upon substituting \eqref{eq:value_tv}, we obtain
\begin{equation}\label{eq:temporal_eq_der}
\partial_u\ln C =-\frac{1}{2} \partial_u\ln\left(\vert \vec{\tau}_v\vert^2+\Lambda^2\right) + \frac{\delta_1 \delta_3}{\Lambda} \frac{1}{\wp(u)+2e_2+\Lambda^2}.
\end{equation}
We define the quantities $A_{1/2}$ so that
\begin{align}
\wp\left(A_1\right) = -2e_2-\Lambda_1^2, &\quad \wp\left(A_2\right) = -2e_2-\Lambda_2^2,\label{eq:wpa} \\
\wp^\prime(A_1)=-2\frac{\delta_1 \delta_3}{\Lambda_1} , &\quad \wp^\prime(A_2)=2\frac{\delta_1 \delta_3}{\Lambda_2} . \label{eq:wppa}
\end{align}
These equations are compatible, since 
\begin{equation}
4\left(\frac{\delta_1 \delta_3}{\Lambda_{1/2}}\right)^2=4\wp\left(A_{1/2}\right)^3-g_2\wp\left(A_{1/2}\right)-g_3 ,
\end{equation}
which is the usual form of the Weierstrass equation, where the moduli $g_2$ and $g_3$ are given by \eqref{eq:elliptic_moduli}. Using \eqref{eq:wpa} and \eqref{eq:wppa}, equation \eqref{eq:temporal_eq_der} assumes the form
\begin{align}
\partial_u\ln C^{\pm 1}&=-\frac{1}{2} \partial_u\ln\left(\wp(u)-\wp(A_1)\right)\mp\frac{1}{2}\frac{\wp^\prime(A_1)}{\wp(u)-\wp(A_1)}, \\
\partial_u\ln C^{\pm 2}&=-\frac{1}{2} \partial_u\ln\left(\wp(u)-\wp(A_2)\right)\pm\frac{1}{2}\frac{\wp^\prime(A_2)}{\wp(u)-\wp(A_2)}.
\end{align}
Thus, the second equation of the auxiliary system is solved by
\begin{align}
C_i^{\pm 1} (u) &= c_i^{\pm 1}\left(\wp(u)-\wp(A_1)\right)^{-\frac{1}{2}} \exp\left(\pm\Phi_1(u)\right),\label{eq:C1pm}\\
C_i^{\pm 2} (u) &= c_i^{\pm 2}\left(\wp(u)-\wp(A_2)\right)^{-\frac{1}{2}} \exp\left(\mp i\Phi_2(u)\right),\label{eq:C2pm}
\end{align}
where $c_i^{\pm 1}$ and $c_i^{\pm 2}$ are constants and
\begin{align}
\Phi_1^\prime(u)&=-\frac{1}{2}\frac{\wp^\prime(A_1)}{\wp(u)-\wp(A_1)},\label{eq:Phi1_prime}\\
\Phi_2^\prime(u)&=\frac{i}{2}\frac{\wp^\prime(A_2)}{\wp(u)-\wp(A_2)}.\label{eq:Phi2_prime}
\end{align}

Equations \eqref{eq:wpa} and \eqref{eq:wppa} are defined so that $A_{1/2}$ possess the property
\begin{equation}
A_{1/2}\vert_{\lambda=0}=a_{1/2}
\end{equation}
which implies that
\begin{equation}
\Phi_{1/2}(u)\vert_{\lambda=0}=\phi_{1/2}(u) .
\end{equation}
The above imply that the quantities $A_{1/2}$ are a natural generalization of the quantities $a_{1/2}$ for the dressed solution, as well as the functions $\Phi_{1/2}$ that appear in the dressed solution are a natural generalization of the functions $\phi_{1/2}$ that appear in the seed solution. Moreover, $\Phi_i$ obey 
\begin{equation}
\bar{\Phi}_{1/2}^\prime\left(u;\bar{\lambda}\right)=\Phi_{1/2}^\prime\left(u;-\lambda\right),
\end{equation}
upon complex conjugation. 

In order to write the solution in a manifestly real form, we introduce the vectors
\begin{align}
E_1=\frac{1}{2}\frac{1}{\sqrt{L_1^2+L_2^2}}\left(V^{+1}-V^{-1}\right),&\quad
E_2=\frac{1}{2}\frac{1}{\sqrt{L_1^2+L_2^2}}\left(V^{+1}+V^{-1}\right),\\
E_3=\frac{1}{2i}\frac{1}{\sqrt{L_1^2+L_2^2}}\left(V^{+2}-V^{-2}\right),&\quad
E_4=\frac{1}{2}\frac{1}{\sqrt{L_1^2+L_2^2}}\left(V^{+2}+V^{-2}\right).
\end{align}
Their explicit expressions are
\begin{align}
E_1=\frac{1}{\sqrt{L^2_1+L^2_2}}\begin{pmatrix}
\sqrt{\wp(u)-\wp(A_1)}\\
\frac{\vec{\tau}_v\times\vec{\kappa}_v}{\sqrt{\wp(u)-\wp(A_1)}}
\end{pmatrix},\qquad E_2=\frac{1}{\sqrt{L^2_1+L^2_2}}\begin{pmatrix}
0\\
\frac{\frac{\delta_1}{L_1}\vec{\tau}_v+L_1\vec{\kappa}_v}{\sqrt{\wp(u)-\wp(A_1)}}
\end{pmatrix},\\
E_3=\frac{1}{\sqrt{L^2_1+L^2_2}}\begin{pmatrix}
\sqrt{\wp(u)-\wp(A_2)}\\
\frac{\vec{\tau}_v\times\vec{\kappa}_v}{\sqrt{\wp(u)-\wp(A_2)}}
\end{pmatrix},\qquad E_4=\frac{1}{\sqrt{L^2_1+L^2_2}}\begin{pmatrix}
0\\
\frac{\frac{\delta_1}{L_2}\vec{\tau}_v-L_2\vec{\kappa}_v}{\sqrt{\wp(u)-\wp(A_2)}}
\end{pmatrix}.
\end{align}

Then, defining
\begin{align}
\mathcal{V}_1&=E_1 \cosh\left(L_1 v+\Phi_1(u)\right) + E_2 \sinh\left(L_1v+\Phi_1(u)\right),\\
\mathcal{V}_2&=E_1 \sinh\left(L_1 v+\Phi_1(u)\right) + E_2 \cosh\left(L_1v+\Phi_1(u)\right),\\
\mathcal{V}_3&=E_3 \cos\left(L_2 v-\Phi_2(u)\right) + E_4 \sin\left(L_2 v-\Phi_2(u)\right),\\
\mathcal{V}_4&=E_3 \sin\left(L_2 v-\Phi_2(u)\right) - E_4 \cos\left(L_2 v-\Phi_2(u)\right),
\end{align}
the solution of the auxiliary system reads
\begin{equation}
\hat{\Psi}=\mathcal{V}\mathcal{C},
\end{equation}
where $\mathcal{V}$ is a matrix, whose columns are $\mathcal{V}_i$ and $\mathcal{C}$ is a constant matrix.

The relation \eqref{eq:psihat_definition} implies that the constraints \eqref{eq:reality_constraint}, \eqref{eq:coset_constraint} and \eqref{eq:subgroup_constraint} for the matrix $\Psi$ translate to
\begin{align}
\bar{\hat{\Psi}}(\bar{\lambda})&=\hat{\Psi}(-\lambda),\label{eq:constraint_psi_hat_bar}\\
J\hat{\Psi}^T(\lambda)J&=\hat{\Psi}^{-1}(\lambda),\label{eq:constraint_psi_hat_inv}\\
J\hat{\Psi}(1/\lambda)J&=\hat{\Psi}(\lambda)m_2(\lambda)\label{eq:constraint_coset},
\end{align}
for the matrix $\hat{\Psi}$\footnote{In general two more constant matrices $m_1$ and $m_3$ should appear in the constraints \eqref{eq:constraint_psi_hat_bar} and \eqref{eq:constraint_psi_hat_inv} (see Section \ref{subsec:constraints}). For simplicity, we set them equal to the identity matrix, without loss of generality.}. Moreover, we remind the reader that the matrix $\hat{\Psi}$ must obey the normalization condition \eqref{eq:Psi_initial_condition}.
We recall that for the special choice of $Y_0$ that we have made, $m_2$ should satisfy
\begin{equation}
m_2(\lambda)J m_2(1/\lambda)J=I\label{eq:constraint_m2}.
\end{equation}
We let the matrix $m_2$ in the constraints unspecified, since this freedom will be required in order to satisfy them. The matrix $\mathcal{V}$ obeys the following relations:
\begin{align}
\mathcal{V}(0)&=-J U^T,\\
\bar{\mathcal{V}}(\bar{\lambda})&=\mathcal{V}(-\lambda),\\
\mathcal{V}^{-1}(\lambda)&=J\mathcal{V}^T(\lambda)J,\\
\mathcal{V}(\lambda)&=-J\mathcal{V}(1/\lambda).
\end{align}
The last one implies that \eqref{eq:constraint_coset} is satisfied for any $\mathcal{C}(\lambda),$ since we can always select
\begin{equation}
m_2(\lambda)=-\mathcal{C}^{-1}(\lambda)J\mathcal{C}(1/\lambda)J,
\end{equation}
so that both \eqref{eq:constraint_coset} and \eqref{eq:constraint_m2} hold true. This means that the non-trivial constraints for the constant matrix $\mathcal C$ are
\begin{align}
\mathcal{C}(0)&=-J,\\
\bar{\mathcal{C}}(\bar{\lambda})&=\mathcal{C}(-\lambda),\\
\mathcal{C}^{-1}(\lambda)&=J\mathcal{C}^T(\lambda)J.
\end{align}
These are trivially satisfied by choosing
\begin{equation}
\mathcal{C}(\lambda)=\mathcal{C}(0)=-J .
\end{equation}
This choice implies that $m_2(\lambda)=-I$. Putting everything together, the solution of the auxiliary system, that satisfies all appropriate constraints, reads
\begin{equation}
\hat{\Psi}=-\mathcal{V}J.
\end{equation}

\subsection{Doubly Dressed Elliptic Minimal Surfaces}
\label{subsec:Doubly_Dressed_Elliptic}
In this section we construct the simplest real dressed elliptic minimal surfaces, using the machinery developed in Sections \ref{sec:Dressing} and \ref{sec:properties}. These are obviously the doubly dressed elliptic minimal surfaces, dressed with the simplest dressing factor, i.e. the one with just a pair of poles lying on the imaginary axis. In everything that follows we drop the indices on $\hat{\Psi}$ that were introduced in the section \ref{subsec:multiple_dressing}. In this section, the symbol $\hat{\Psi}$ always refers to the solution of the auxiliary system that corresponds to the elliptic minimal surfaces, which was derived in Section \ref{subsec:auxiliary_solution_elliptic}. In this case the matrix $\mathcal{U}$ of \eqref{eq:def_V_tilde} coincides with $U$, thus
\begin{equation}
\tilde{V}_k=\hat V_k=\hat{\Psi}\left(i \mu_k\right)p_k
\end{equation}
Equation \eqref{eq:Yk_2} implies that the temporal and spatial components of  $\hat{Y}_2$ are
\begin{align}
\hat{Y}_2^0&=\left(1-\frac{\left(1+\mu_{1}^{-1}\mu_{2}^{-1}\right)\left(1+\mu_{1}\mu_{2}\right)}{2X}\right)\\
\vec{\hat{Y}}_2&=\frac{1}{2X}\frac{1+\mu_{1}\mu_{2}}{\mu_{2}-\mu_{1}}\left[\left(\mu_2+\mu_2^{-1}\right)\hat{n}_2-\left(\mu_{1}+\mu_{1}^{-1}\right)\hat{n}_1\right],
\end{align}
where $\hat{n}_1$ and $\hat{n}_2$ are unit norm vectors, which are given by \eqref{eq:nhat} and $X$ is given by \eqref{eq:def_X_nhat}.

The constant vectors $p_k$ can be parametrized as
\begin{equation}\label{eq:projector}
p_k=\begin{pmatrix}
\cosh\theta^0_k \\ \sinh\theta^0_k \\ \cos\phi^0_k\\ \sin\phi^0_k
\end{pmatrix},
\end{equation}
so that they are manifestly null\footnote{Since $p_k$ is null it can be parametrized as $p_k^T=a\left(1,\cos\theta,\sin\theta\cos\phi,\sin\theta\sin\phi\right)$, or $p_k^T=a\sin\theta\left(\frac{1}{\sin\theta},\tan\theta,\cos\phi,\sin\phi\right)$. Taking into account the fact that equation \eqref{eq:Yk_2} is homogeneous in $p_k$ we can drop the overall factor and define $\cosh u=\frac{1}{\sin\theta}$ and $\sinh u=\tan\theta$. Thus, \eqref{eq:projector} is the most general form of $p_k$.}. Then, the temporal component of $V_k$ is
\begin{equation}
\hat V^0_k=\frac{1}{\sqrt{L_{1,k}^2+L_{2,k}^2}}\left(
\sqrt{\wp(u)-\wp(A_{1,k})}\cosh\left(\Omega_{1,k}\right)-\sqrt{\wp(u)-\wp(A_{2,k})}\cos\left(\Omega_{2,k}\right)
\right),
\end{equation}
while the spatial components are
\begin{multline}
\vec{\hat{V}}_k=\frac{1}{\sqrt{L_{1,k}^2+L_{2,k}^2}}\left[\vec{\tau}_v\times\vec{\kappa}_{v,k}\left(\frac{\cosh\left(\Omega_{1,k}\right)}{\sqrt{\wp(u)-\wp(A_{1,k})}}-\frac{\cos\left(\Omega_{2,k}\right)}{\sqrt{\wp(u)-\wp(A_{2,k})}}\right) \right.\\
+\left(\frac{\delta_{1,k}}{L_{1,k}}\vec{\tau}_v+L_{1,k}\vec{\kappa}_{v,k}\right)\frac{\sinh\left(\Omega_{1,k}\right)}{\sqrt{\wp(u)-\wp(A_{1,k})}}\\-\left.\left(\frac{\delta_{1,k}}{L_{2,k}}\vec{\tau}_v-L_{2,k}\vec{\kappa}_{v,k}\right)\frac{\sin\left(\Omega_{2,k}\right)}{\sqrt{\wp(u)-\wp(A_{2,k})}}\right].
\end{multline}
We use the shorthand notation
\begin{align}
\Omega_{1,k}&=L_{1,k} v +\Phi_1(u;A_{1,k}) - \theta^0_k,\\
\Omega_{2,k}&=L_{2,k} v -\Phi_2(u;A_{2,k}) - \phi^0_k.
\end{align}
The parameters of the solution of the auxiliary system satisfy the equations
\begin{align}
\wp(A_{1,k})&=-\frac{1}{12}E-\frac{1}{4}\sqrt{E^2+16\left(\delta_{1,k}\right)^2},\\
\wp(A_{2,k})&=-\frac{1}{12}E+\frac{1}{4}\sqrt{E^2+16\left(\delta_{1,k}\right)^2}.
\end{align}
Since $\left(\delta_{1,k}\right)^2\leq 1/4,$ in view of \eqref{eq:delta_13_sq}, we obtain $e_3\leq \wp(A_{1,k})\leq e_2$ and $e_2\leq \wp(A_{2,k})\leq e_1$, similarly to the inequalities \eqref{eq:elliptic_range} obeyed by the analogous quantities $a_{1,2}$ of the seed solution. This is expected from the band structure of the $n=1$ \Lame potential. These constraints ensure that the \Lame phases defined in \eqref{eq:Phi1_prime} and \eqref{eq:Phi2_prime} are real.

Finally, we rotate the vector $\hat Y_2$ back to the unhatted coordinate system of the enhanced space $Y$, through equation \eqref{eq:rotation}, and we obtain the following expression for the dressed solution
\begin{equation}
Y=\begin{pmatrix}
\left(F_1\hat{Y}_2^0+F_2\hat{Y}_2^2\right) \cosh\left(\ell_1 v+\phi_1(u)\right) + \hat{Y}_2^1 \sinh\left(\ell_1 v+\phi_1(u)\right)\\
\left(F_1\hat{Y}_2^0+F_2\hat{Y}_2^2\right) \sinh\left(\ell_1 v+\phi_1(u)\right) + \hat{Y}_2^1 \cosh\left(\ell_1 v+\phi_1(u)\right)\\
\left(F_2\hat{Y}_2^0+F_1\hat{Y}_2^2\right) \cos\left(\ell_2 v-\phi_2(u)\right) - \hat{Y}_2^3 \sin\left(\ell_2 v-\phi_2(u)\right)\\
\left(F_2\hat{Y}_2^0+F_1\hat{Y}_2^2\right) \sin\left(\ell_2 v-\phi_2(u)\right) + \hat{Y}_2^3 \cos\left(\ell_2 v-\phi_2(u)\right)
\end{pmatrix} .
\end{equation}
After a tedious calculation one can show that $u=2n\omega_1,$ where $n\in\mathbb{N},$ does not correspond to the AdS boundary, unlike the elliptic precursor of the dressed solution. The boundary of the dressed minimal surface is determined by the equation
\begin{equation}
X=0.
\end{equation}

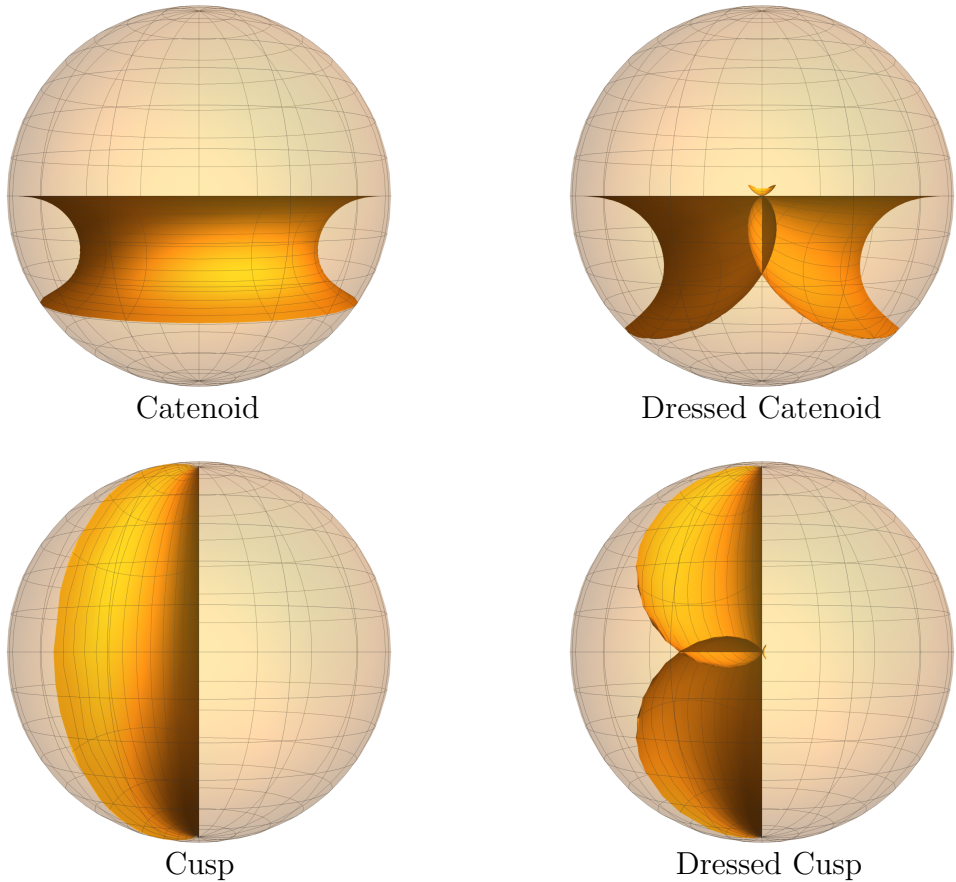
\begin{figure}[ht]
\vspace{10pt}
\begin{center}
\begin{picture}(98,85)
\put(1,40){\includegraphics[width = 0.45\textwidth]{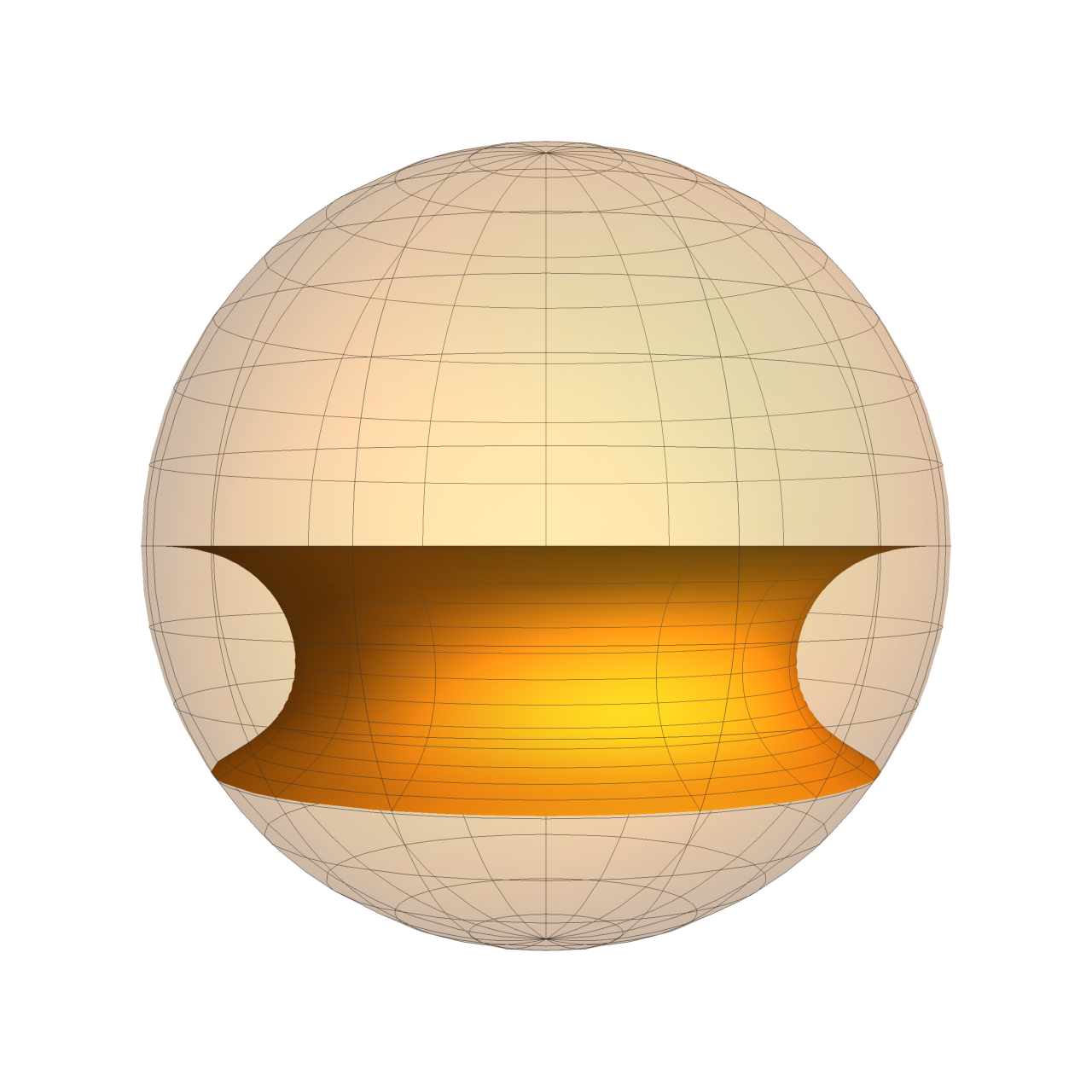}}
\put(50,40){\includegraphics[width = 0.45\textwidth]{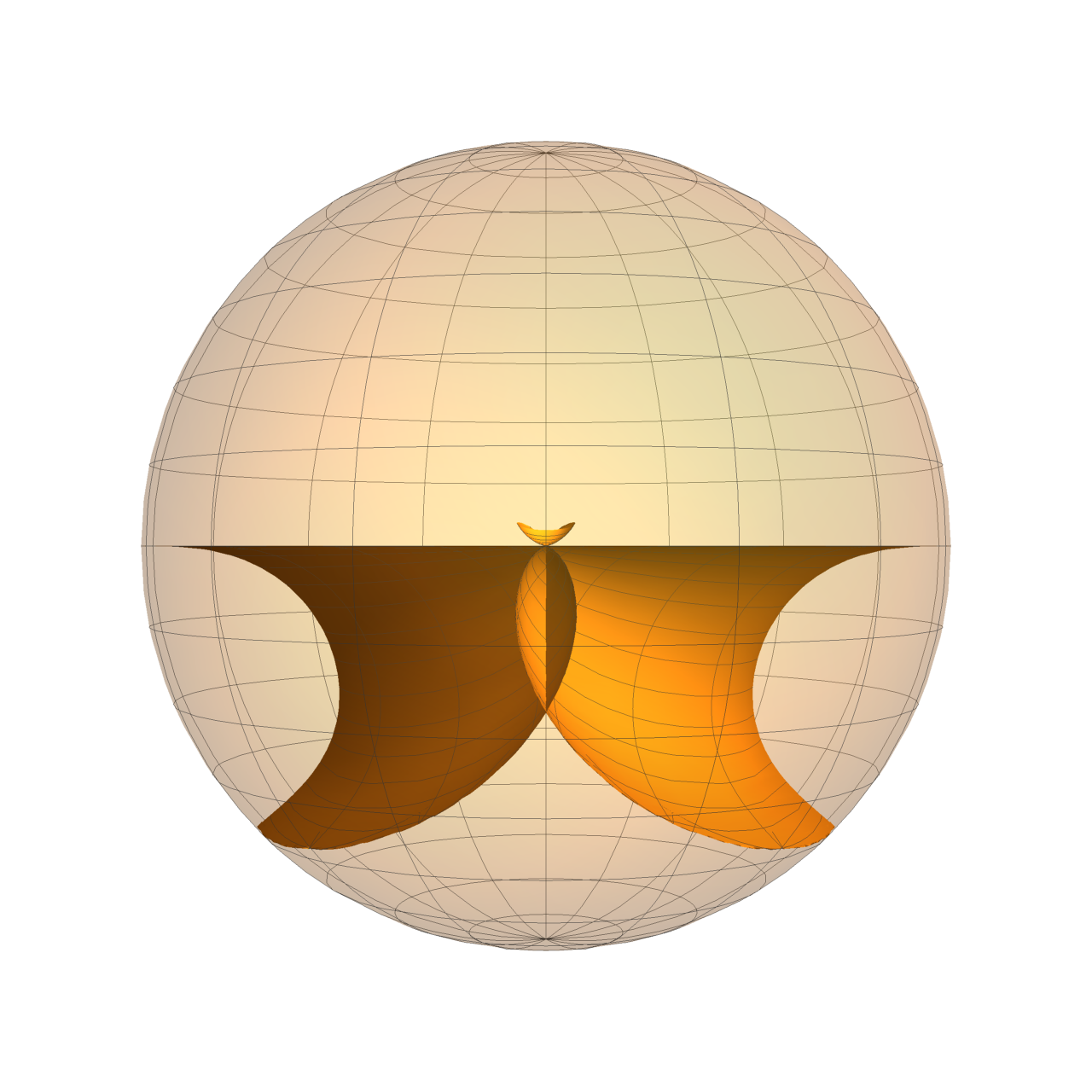}}
\put(1,0){\includegraphics[width = 0.45\textwidth]{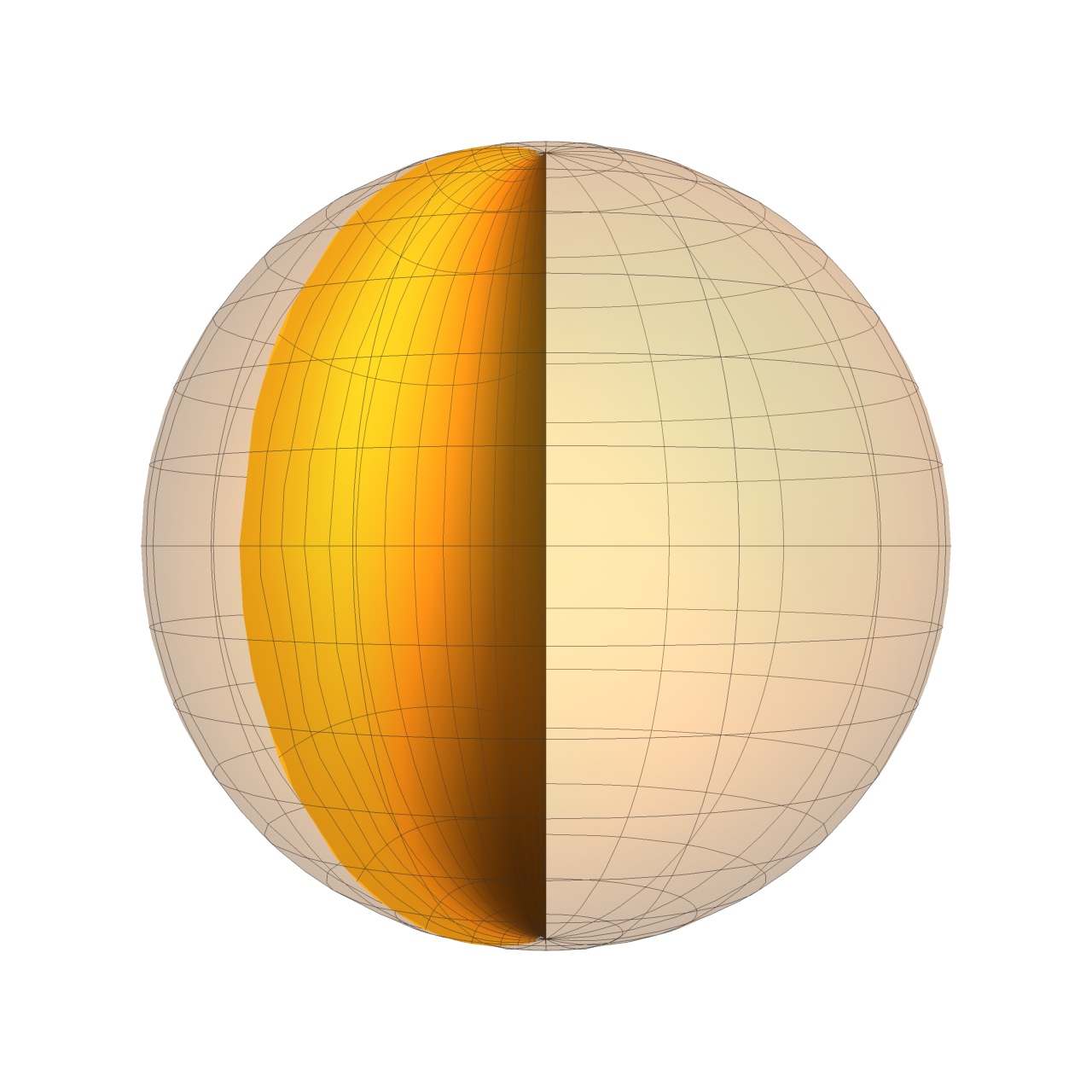}}
\put(50,0){\includegraphics[width = 0.45\textwidth]{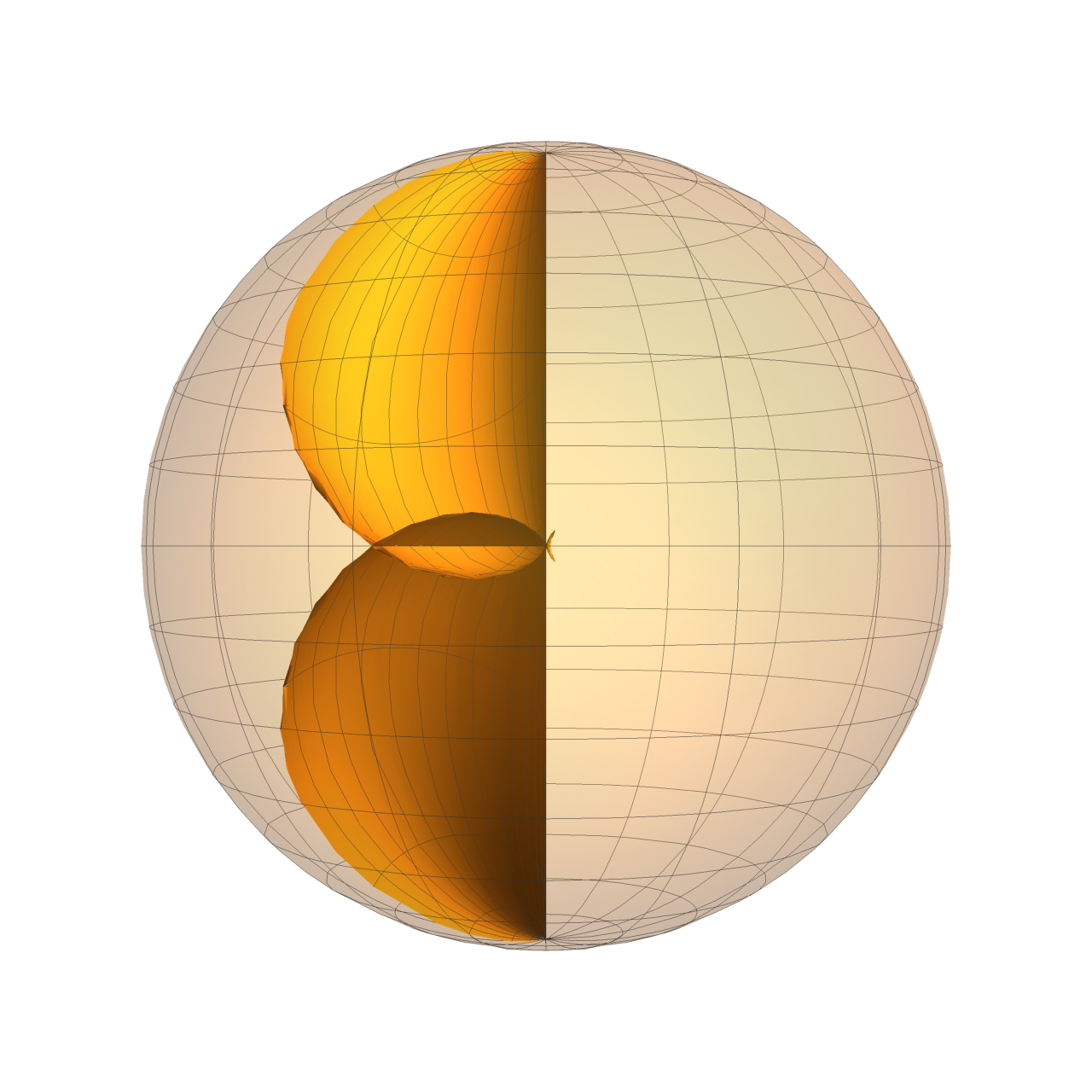}}
\put(18,43){Catenoid}
\put(62,43){Dressed Catenoid}
\put(20.5,3){Cusp}
\put(65,3){Dressed Cusp}
\end{picture}
\end{center}
\vspace{-15pt}
\caption{Two representative dressed elliptic minimal surfaces and their seeds in global coordinates. In the plot the radial coordinate corresponds to the tortoise coordinate $r^*=\arctan r$, so that the surface $r^*=\pi/2$ is the AdS boundary.}
\label{fig:plots}
\end{figure}

\begin{figure}[ht]
\vspace{10pt}
\begin{center}
\begin{picture}(98,45)
\put(1,0){\includegraphics[width = 0.45\textwidth]{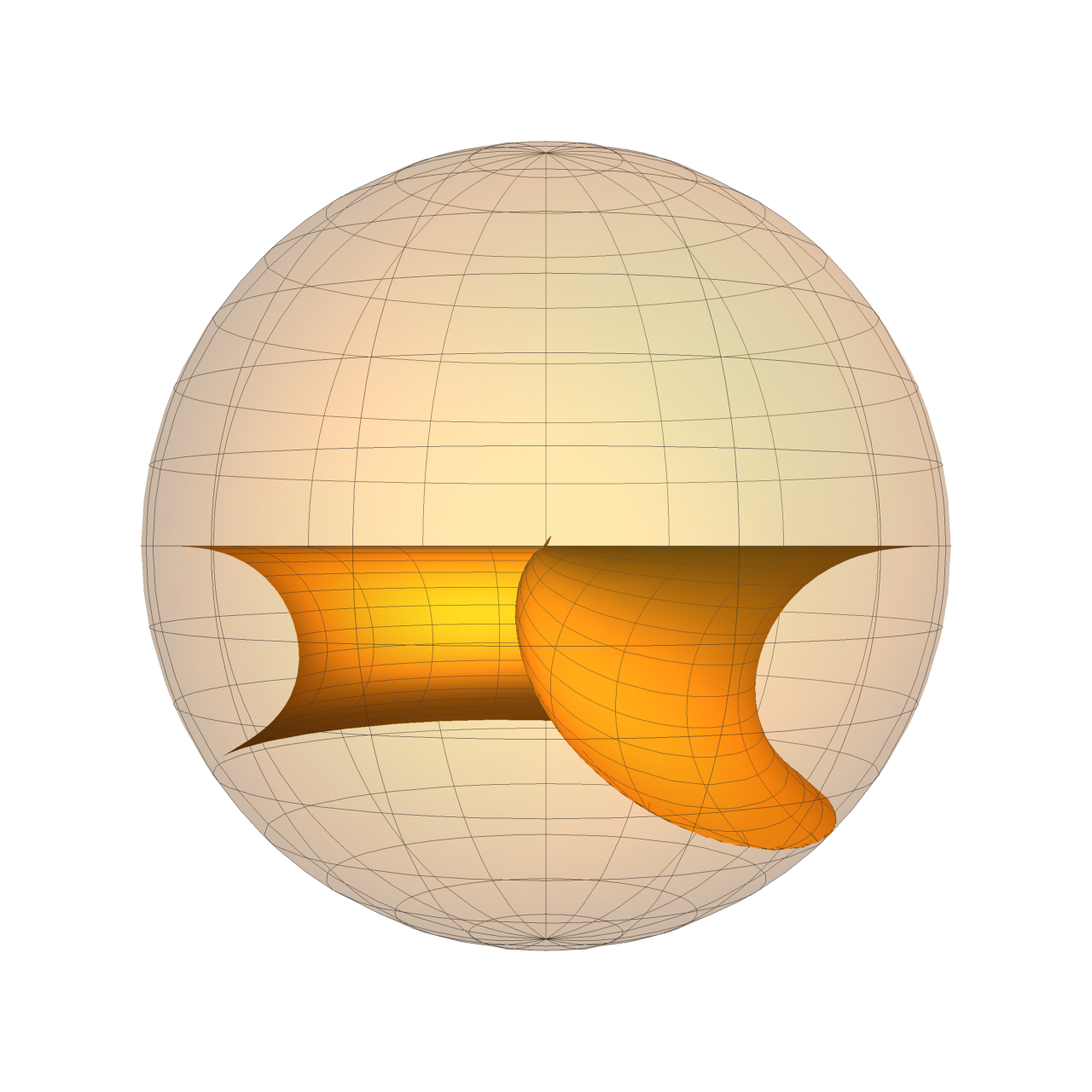}}
\put(50,0){\includegraphics[width = 0.45\textwidth]{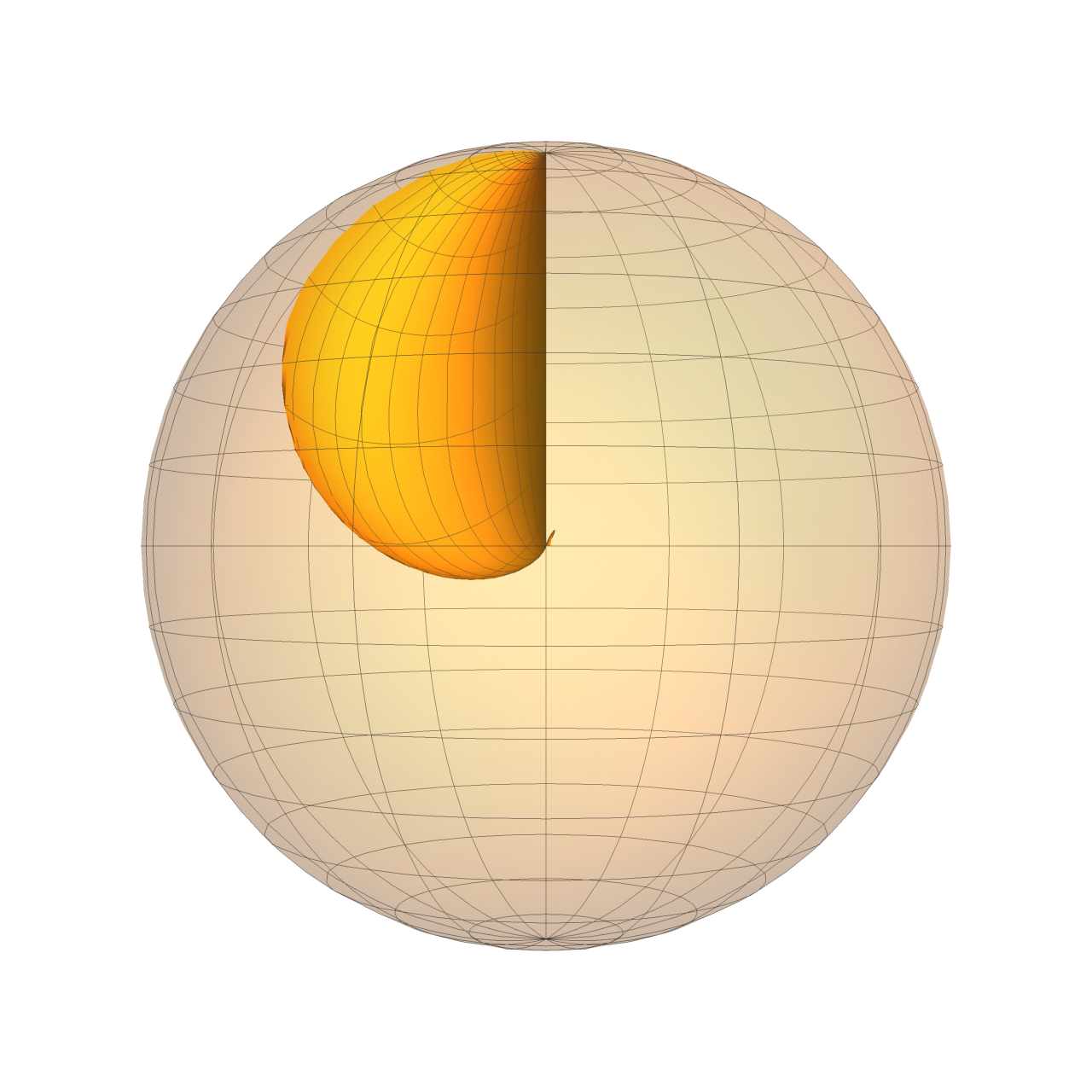}}
\put(13.5,3){Dressed Catenoid}
\put(65,3){Dressed Cusp}
\end{picture}
\end{center}
\vspace{-15pt}
\caption{The dressed catenoid and cusp, plotted in an appropriate subset of the world-sheet parameters of their seeds, so that they are anchored at the boundary, yet the do not possess self-intersections.}
\label{fig:plots_half}
\end{figure}

In order to visualize the effect of the dressing transformation on an elliptic minimal surface, we present two indicative examples in figure \ref{fig:plots}.
These examples employ a catenoid and a cusp as seed minimal surfaces. It is evident that the boundary, which is the corresponding entangling curve, is altered in a non-trivial manner. The effect of the dressing transformation on the minimal surfaces is similar to the one on string solutions \cite{Katsinis:2019oox}. The deformation of the surfaces is localized in a specific region, whereas asymptotically the dressed solution recovers the form of its seed. Intuitively, the deformed region corresponds to the location of the solitons inserted by the dressing transformation in the Pohlmeyer counterpart. It appears that the dressed elliptic minimal surfaces have self-intersections in the aforementioned region, which are analogous to the loops that appear in dressed elliptic strings. The self-intersections imply that these surfaces are not the globally preferred ones that correspond to the specific boundary conditions. Nevertheless, one can restrict the world-sheet parameters in appropriate regions, so that the surface is still anchored at the boundary and does not have any self-intersections, see figure \ref{fig:plots_half}.

\section{Discussion}
\label{sec:discussion}

We presented the construction of the dressed static elliptic minimal surfaces in AdS$_4$. The auxiliary system for a general elliptic seed solution was solved, and, subsequently, an arbitrary number of dressing transformations was applied. This led to a recursive construction of NLSM solutions out of the initial elliptic solutions. For this, the simplest possible dressing factor was used, namely, the one containing two poles on the imaginary axis. We showed, that this particular type of dressing factor acts as a boost with superluminal velocity on the seed solution. 

It turned out that only an even number of dressing transformations results in new real solutions of the NLSM in H$^3$, that correspond to static minimal surfaces in AdS$_4$. The application of an odd number of dressing transformations leads to purely imaginary solutions in H$^3$, which correspond to real solutions of the NLSM on dS$_3$. The fact that the dressing method connects solutions of the Euclidean NLSM on H$^3$ to solutions of the Euclidean NLSM on dS$_3$ and vice versa is analogous to B\"acklund transformations, which connect solutions of different differential equations. 

Furthermore, we obtained a recursive relation between the surface element on the seed minimal surface and the one on the dressed minimal surface, which emerges after a double dressing transformation. Unfortunately, we could not do more than that in the direction of computing the area of the dressed minimal surface. Since we were not able to determine the boundary region of the minimal surface, we do not know the domain of integration of the surface element. These difficulties originate from the inherent complexity of the static elliptic minimal surfaces, which are expressed in terms of Weierstrass elliptic functions. Clearly, in view of the AdS/CFT correspondence, it would be interesting to overcome the aforementioned difficulties and to compute the area of the dressed minimal surfaces and how this is altered by the dressing.

Naively, it seems that the existence of self-intersections is an inherent characteristic of the dressed elliptic minimal surfaces. The strong sub-additivity of holographic entanglement entropy suggests that these minimal surfaces do not correspond to the globally minimal ones. However, by restricting the world-sheet parameters in appropriate regions, this problem can be resolved. Therefore, the presented minimal surfaces can find applications in the context of holographic entanglement entropy. The alteration of the entangling curve by the dressing transformation is complicated. It would be interesting to investigate whether one could perform a dressing transformation that leaves the boundary region intact. In such a case the dressing transformation could probe directly the stability of the seed minimal surface in the same fashion as it does for elliptic string solutions\cite{Katsinis:2019sdo}.

A possible future extension of this work is to find the Pohlmeyer counterpart of the dressed solution and relate it to the Pohlmeyer counterpart of the seed solution. The NLSM on H$^3$ can be mapped via Pohlmeyer reduction to the $\cosh$-Gordon equation. A parallel construction of the elliptic solutions on both sides of this mapping was presented in \cite{Pastras:2016vqu}. The establishment of an analogous correspondence for the dressed minimal surfaces presents a certain interest. According to a similar analysis that was performed for the NLSM on S$^2$ in \cite{Katsinis:2018ewd}, it is expected that the Pohlmeyer counterpart of the dressed solution will be connected through a finite number of B\"acklund transformations with the Pohlmeyer counterpart of the seed solution. The $\cosh$-Gordon equation lacks a vacuum, and, thus, the simplest solutions to be used as seed for the application of B\"acklund transformations, are the elliptic ones. Consequently, the Pohlmeyer counterparts of the dressed solutions should be some of the simplest kink-like solutions of the $\cosh$-Gordon equation.

An alternative approach for the construction of dressed minimal surfaces is the application of a single dressing transformation with the simplest dressing factor on imaginary seeds corresponding to elliptic solutions of the Euclidean NLSM defined on dS$_3$. For this purpose, the latter should be first constructed via methods similar to those in \cite{Pastras:2016vqu}.

\subsection*{Acknowledgements}
The research of D.M., I.M. and G.P. has received funding from the Hellenic Foundation for Research and Innovation (HFRI) and the General Secretariat for Research and Technology (GSRT), in the framework of the ``First Post-doctoral researchers support'', under grant agreement No 2595. The research of D.K. is co-financed by Greece and the European Union (European Social Fund- ESF) through the Operational Programme ``Human Resources Development, Education and Lifelong Learning'' in the context of the project ``Strengthening Human Resources Research Potential via Doctorate Research'' (MIS-5000432), implemented by the State Scholarships Foundation (IKY). The authors would like to thank M. Axenides and E. Floratos for useful discussions.

\appendix

\section{The Equations of Motion and the Virasoro Constraints}
\label{sec:appendix}
In order to verify that the dressed minimal surface $Y_k$, which is given by \eqref{eq:Y_k}, satisfies the Virasoro constraints, we use the auxiliary system \eqref{eq:psi_k_auxliary}. Projecting it in the direction of the vector $p_k$ yields
\begin{equation}
\partial_\pm W_k=\frac{1}{1\pm i \mu_k}\left(\partial_\pm g_{k-1}\right)g_{k-1}^{-1}W_k.
\end{equation} 
Taking into account the mapping \eqref{eq:gk_mapping}, after some algebra we obtain
\begin{equation}\label{eq:W_der}
\partial_\pm W_k=\frac{2}{1\pm i \mu_k}J\left[\left(W_k^T\partial_\pm Y_{k-1}\right)Y_{k-1} -\left(W_k^T Y_{k-1}\right)\partial_\pm Y_{k-1}\right].
\end{equation}
In addition, since  $Y_{k-1}^TJY_{k-1}=-1$, we obtain
\begin{equation}\label{eq:WY_der}
\partial_\pm \left(W_k^T Y_{k-1}\right)=-\frac{1\mp i \mu_k}{1\pm i \mu_k}W_k^T \partial_\pm Y_{k-1}.
\end{equation}
Putting everything together, the derivatives of $Y_k$ assume the form
\begin{equation}\label{eq:Yk_der}
\partial_\pm Y_k=i\left[\pm i \partial_\pm Y_{k-1}+\frac{1\mp i\mu_k}{\mu_k}\frac{W_k^T \partial_\pm Y_{k-1}}{W_k^T Y_{k-1}}Y_{k-1}+\frac{\left(1\mp i\mu_k\right)^2}{2\mu_k}\frac{W_k^T \partial_\pm Y_{k-1}}{\left(W_k^T Y_{k-1}\right)^2}JW_k\right].
\end{equation}
Then, it is a matter of algebra to show that
\begin{equation}
\left(\partial_\pm Y_k\right)^T J \left(\partial_\pm Y_k\right) = \left(\partial_\pm Y_{k-1}\right)^T J \left(\partial_\pm Y_{k-1} \right),
\end{equation}
thus, the solution $Y_k$ satisfies the Virasoro constraints, as long as its seed $Y_{k-1}$ does so.

Similarly, one can show that the surface element transforms as
\begin{equation}
\left(\partial_+ Y_k\right)^T J \left(\partial_- Y_k\right) = -\left(\partial_+ Y_{k-1}\right)^T J \left(\partial_- Y_{k-1}\right) + 2\frac{\left(W_k^T \partial_+ Y_{k-1}\right)\left(W_k^T \partial_- Y_{k-1}\right)}{\left(W_k^T Y_{k-1}\right)^2} .
\end{equation}
Taking into account \eqref{eq:WY_der}, we obtain
\begin{equation}\label{eq:pohl_addition}
\left(\partial_+ Y_k\right)^T J \left(\partial_- Y_k\right) = -\left(\partial_+ Y_{k-1}\right)^T J \left(\partial_- Y_{k-1}\right) + 2\frac{\partial_+ \left(W_k^T Y_{k-1}\right)\partial_-\left(W_k^T Y_{k-1}\right)}{\left(W_k^T Y_{k-1}\right)^2}.
\end{equation}
Using \eqref{eq:W_der} and \eqref{eq:WY_der} it is easy to show that
\begin{equation}\label{eq:WY_der2}
\partial_+\partial_-\left(W_k^T Y_{k-1}\right)=\left(W_k^T Y_{k-1}\right)\left(\partial_+ Y_{k-1}\right)^T J \left(\partial_- Y_{k-1}\right).
\end{equation}

In order to show that the equations of motion of $Y_k$ are satisfied, we substitute \eqref{eq:Yk_der} into \eqref{eq:WY_der}, so that the latter assumes the form
\begin{equation}
\partial_\pm Y_k=\mp\left(\partial_\pm Y_{k-1}-\frac{\partial_\pm\left(W_k^T Y_{k-1}\right)}{W_k^T Y_{k-1}}Y_{k-1}\right)-\frac{\partial_\pm\left(W_k^T Y_{k-1}\right)}{W_k^T Y_{k-1}}Y_{k}.
\end{equation}
Then, with the aid of \eqref{eq:WY_der2} it is a matter of algebra to show that
\begin{multline}
\partial_+\partial_-Y_k+\left[\left(\partial_+ Y_{k-1}\right)^T J \left(\partial_- Y_{k-1}\right) - 2\frac{\partial_+ \left(W_k^T Y_{k-1}\right)\partial_-\left(W_k^T Y_{k-1}\right)}{\left(W_k^T Y_{k-1}\right)^2}\right]Y_k \\
= - \partial_+\partial_-Y_{k-1}+\left[\left(\partial_+ Y_{k-1}\right)^T J \left(\partial_- Y_{k-1}\right)\right]Y_{k-1},
\end{multline}
which in view of \eqref{eq:pohl_addition}, proves that the vector $Y_k$ satisfies the equations of motion, as long as the vector $Y_{k-1}$ does so.

\end{document}